\documentclass[%
 reprint,
 amsmath,amssymb,
 aps,
]{revtex4-2}

\usepackage{graphicx}
\usepackage{dcolumn}
\usepackage{bm}
\usepackage{color, soul}
\usepackage{float}
\usepackage{comment}
\usepackage[font=small]{caption, subcaption}
\usepackage{fancyhdr}
\usepackage{bbm}
\usepackage{caption}
\usepackage{ragged2e}
\usepackage{tikz}
\usepackage{dsfont}
\usepackage{url}

\newcommand{\R}{\mathbb{R}}
\newcommand{\bfx}{{\bf x}}
\newcommand{\Pceil}{P^{\text{ceil}}}
\newcommand{\bfP}{{\bf P}}
\newcommand{\inter}{\text{int}}

\begin{document}
\raggedbottom
\preprint{APS/123-QED}

\title{A dynamical system model of gentrification:\\Exploring a simple rent control strategy}

\author{Jonathan D. Shaw}%
\email{jonathan.shaw@colorado.edu}
\author{Juan G. Restrepo}%
 \email{juanga@colorado.edu}
 \author{Nancy Rodr\'{i}guez}%
 \email{rodrign@colorado.edu}
\affiliation{Department of Applied Mathematics, University of Colorado, Boulder, CO 80309, USA}%

\date{\today}

\begin{abstract}
Motivated by the need to understand the factors driving gentrification, we introduce and analyze two simple dynamical systems that model the interplay between three potential drivers of the phenomenon. The constructed systems are based on the assumption that three canonical drivers exist: a subpopulation that increases the desirability of a neighborhood, the desirability of a neighborhood, and the average price of real estate in a neighborhood. The second model modifies the first and implements a simple rent control scheme. For both models, we investigate the linear stability of equilibria and numerically determine the characteristics of oscillatory solutions as a function of system parameters. 
Introducing a rent control scheme stabilizes the system, in the sense that the parameter regime under which solutions approach equilibrium is expanded. However, oscillatory time series generated by the rent control model are generally more disorganized than those generated by the non-rent control model; in fact, long-term transient chaos was observed under certain conditions in the rent control case. Our results illustrate that even simple models of urban gentrification can lead to complex temporal behavior. 
\end{abstract}

\maketitle

\section{Introduction}
\label{sec:introduction}
The evolution of urban areas plays a significant role in economic and policy planning \cite{stein2019}; predicting how urban areas will develop has proved challenging for many city leaders and planners.  A major issue facing city planners around the United States is gentrification, the process of a demographic shift in a community.  One factor believed to play a role in gentrification is the influx of artists into a neighborhood.  A classic example is the gentrification of Lower Manhattan beginning in the 1950s. We are motivated by this example which will be discussed in more detail in section \ref{sec:motiv}.  

This paper aims to model the interaction between factors such as real estate prices, neighborhood desirability, and the urban migration patterns of specific population segments. Our focus is on the demographics of artists, who are typically viewed as a vulnerable population. Artists' migration can influence a broader demographic known as the creative class, encompassing individuals in art, media, and design, as well as knowledge-based occupations like healthcare, education, finance, business, and the legal profession \cite{florida2003cities}. The rise of the creative class in a region tends to impact average real estate prices. While demographic shifts are intricate, our goal is to present and analyze a parsimonious model that captures key aspects of the gentrification process and a basic rent control strategy. We observe that even a simple model can yield complex and nuanced behaviors, offering insights into the pros and cons of rent-control policies.      

In this study, we first introduce a simple three-component model that produces highly intricate behavior. Notably, we investigate parameter regions that result in the continuous displacement of the artist population. Subsequently, we introduce a basic rent control strategy to examine its impact on displacement. On one hand, we observe an expansion in the size of parameter regions that mitigate the displacement of the artist population. On the other hand, we also identify parameter regions leading to transient chaos, a phenomenon absent in the non-rent control model. This suggests that while some regions may benefit from rent control strategies, others may experience challenges, contributing to the contentious debate between proponents and opponents of rent control policies.

\subsection{Motivation}\label{sec:motiv}
The concept of gentrification was coined by British social scientist Ruth Glass in 1964 in her book {\it London: Aspects of Change} and has been controversial since then \cite{Osman2016}.  According to The Encyclopedia of Housing \cite{Smith1998}, gentrification is the process by which central urban neighborhoods that have undergone disinvestment and economic decline experience a reversal, reinvestment, and the in-migration of a relatively well-off, middle and upper-middle-class population.  In the earlier history of the United States, gentrification did not play a significant role in demographic shifts.  However, since the 1970s it has played a significant role in shaping cities like Seattle, San Francisco, New York, Boston, and Washington DC.  
While in theory gentrification sounds like a desirable outcome for a community, in practice the rise in prices can lead to the displacement of the neighborhood's original inhabitants or cultural displacement \cite{betancur2011}. The mixed outcomes of gentrification have aroused much interest in city planners, community advocates, and the research community. 

In his 1984 essay “The Fine Art of Gentrification,” Moskowitz discusses how artists have contributed to gentrification.  Moskowitz considered two factors to be influential in this process: first, the segregated white community living in single homes with little community and diversity and, second, the commercialization of art.  In this process, artists move into neighborhoods that offer diversity and a low cost of living.  An influx of artists then can begin to change the neighborhood's fabric through their work. 
These changes can also lead to an increase in industry investment, government tax breaks, and grants for real estate, which in turn increase the average price of real estate. Unfortunately, in many cases, the average price of real estate increases to the point that the artists can no longer afford to continue living in the neighborhoods they helped gentrify.    

A classic example is the gentrification of Lower Manhattan which began in the 1950s when artists saw the low rents and open space available in Manhattan as an opportunity.  From the 1960s to the 1980s, artists moved in as illegal residents to lofts and helped transform the decaying industrialized area into an art district famously known as SoHo \cite{Shkuda2016}.  The artists fought to get the lofts zoned for residential use and in 1971 were allowed to apply for certificates to legally reside in the lofts \cite{Winchell2019}.  Developers also saw the opportunity in SoHo, and soon it became unaffordable to the people who helped SoHo become a desirable art district.  
In the 1970s and 1980s, the city certified hundreds of artists to live in SoHo every year, but by 2020, it certified only four \cite{Hughes2021}.

Gentrification is a more complex process than artists moving into a neighborhood.  An inability to study this problem experimentally and the potential for a wide variety of motivating factors complicate any deep understanding of the issue. Hamnett (1991) suggested three main drivers for gentrification -- the existence of middle-class potential gentrifiers, an availability of urban housing, and a tendency among these potential gentrifiers to prefer to live in an urban setting \cite{hamnett1991blind}. Other proposed drivers of gentrification include falling crime rates in inner-city neighborhoods \cite{dastrup2016linking}, demanding work schedules and lack of free time among the young middle class \cite{edlund2015bright}, and proximity to social amenities, such as coffee shops, beer gardens, bike shares, gyms, and restaurants \cite{jeong2021does}.  Future work will involve adding nuances to the model.  For example, we will explore the dynamics of other subpopulations that either influence or are affected by the demographic shift process.  Importantly, we will consider the dynamics of low-income groups, which are especially at risk as well as high-income groups, which also highly influence the increase in the price of real estate.  

\subsection{Prior research}

\subsubsection{Mathematical models of gentrification}

It is fair to say that the majority of theoretical analysis of gentrification has focused on binary divisions -- blacks and whites, flows of capital and flows of people, macro-forces of capital accumulation -- concentrating on subsets of the potential dynamics involved in gentrification \cite{benard2007wealth,zuk2015gentrification}. Typically, modeling gentrification involves agent-based models that allow virtual simulation instead of experiments.  An example of this is the seminal {\it Schelling model} of residential segregation, which uses an agent-based model inhabiting an eight by eight lattice with two classes of agents to represent an arbitrary binary social division \cite{schelling1971dynamic,schelling2006micromotives}. The Schelling model found that segregation was rampant even when agents were willing to inhabit neighborhoods that consisted of up to two-thirds of the other group \cite{schelling1971dynamic,schelling2006micromotives}.

\begin{figure}[t]
    \centering
    \includegraphics[width = 0.4\textwidth]{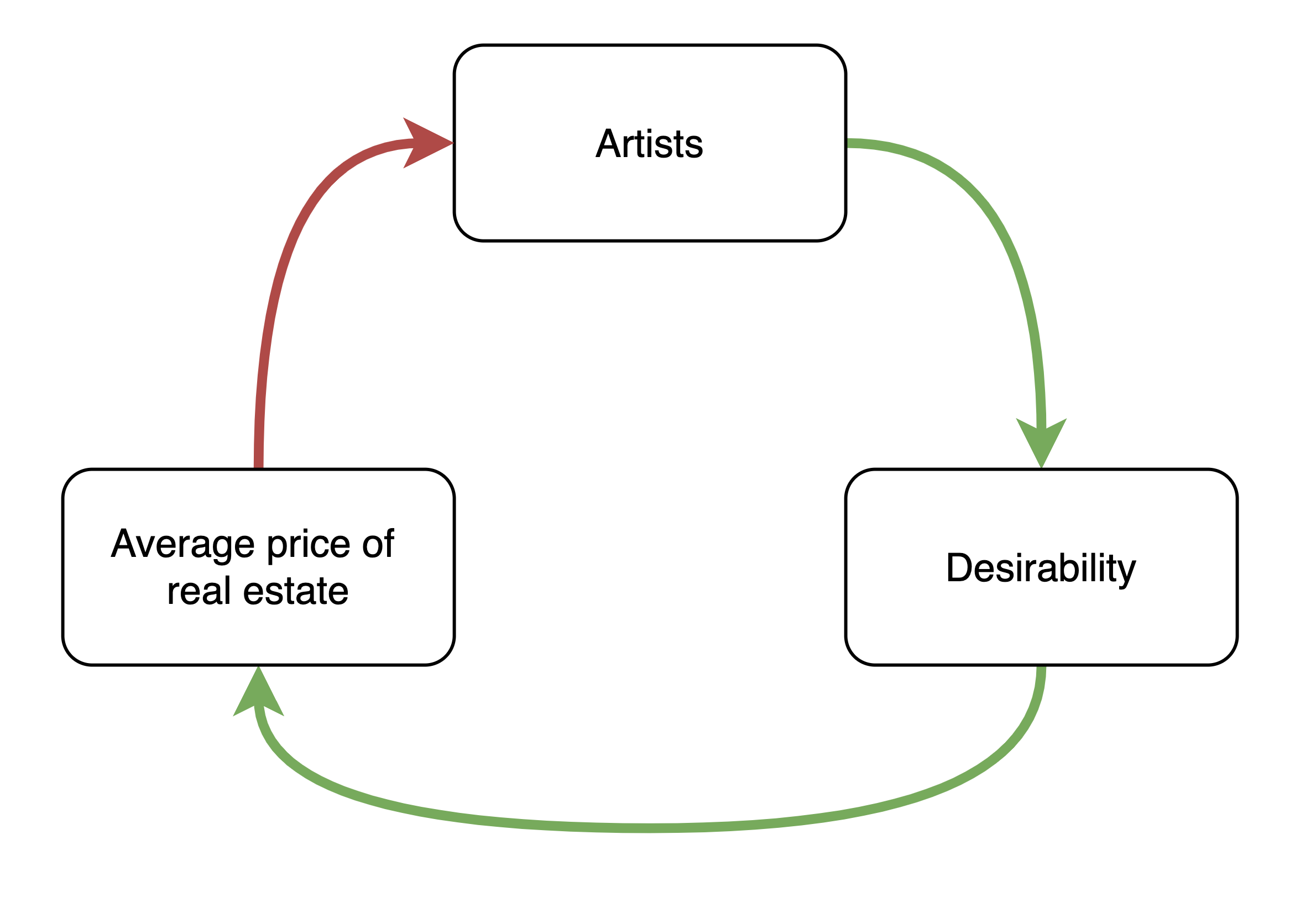}
    \caption{\justifying{Flowchart illustrating the interplay between artists, the desirability of a neighborhood, and the average price of real estate of a neighborhood. Green arrows indicate positive feedback, while the red arrow indicates negative feedback.}}
    \label{fig:fc}
\end{figure}

Extensions of the Schelling model to examine a variety of issues related to residential segregation and gentrification focused on similar agent-based approaches \cite{laurie2003role,pancs2007schelling,zhang2004residential,bruch2006neighborhood}. While agent-based models can provide a detailed and realistic representation of certain systems, they can be challenging to analyze due to the inherent complexity arising from the interactions of numerous agents with potentially heterogeneous behaviors.
On the other hand, continuum models abstract away individual-level interactions and behaviors and instead focus on the overall trends and patterns in a system. Due to their analytical nature, they are often amenable to various mathematical tools and techniques, such as stability analysis, equilibrium analysis, and perturbation analysis. This makes it easier to derive insights and make predictions about the system's behavior.  In this direction, the authors in \cite{Hassan2019a} used partial differential equation systems to model an amenities-based gentrification theory.    

\subsubsection{The paper-rock-scissors analogue}
There is an intransitive relationship between the artist populations, the desirability of a neighborhood, and a neighborhood's average price of real estate.  Specifically, the real estate price of a neighborhood affects the living choices of artists: if the price is high then the density of artists is low. In turn, the 
arrival of artists in a neighborhood increases its desirability.  Finally, an increase in desirability leads to an increase in the average price of a neighborhood. These relationships are illustrated in Figure \ref{fig:fc}, where positive influences are indicated with green arrows and negative influences with red arrows. 

Mathematically, this is analogous to intransitive competition in ecology, where there is no single dominant competitor.  The first mathematical model for intransitive competition was proposed by May and Leonard in \cite{May1975}.  Further mathematical analysis of these types of systems can be found in \cite{Gallien2017,Soliveres2018}.  The intransitive competition affects diversity as it allows for the co-existence of species, which is difficult to obtain with transitive competition models.  

\subsection{Rent controls}\label{sec:rc}

As housing prices in many major cities become out of reach for the majority of Americans, politicians and city planners have looked to rent control policies as a potential solution.  Rent control policies were introduced in the United States and Europe in response to World War II \cite{basu2000}.  In its first generation, rent control imposed strict limits on rent and other costs as part of the war effort, and many of these policies continued in place after the war until the deregulation period of the late 1940s and early 1950s \cite{arnott1995}. 
The decoupling of rents from the markets that occur as a consequence of different rent control policies has been a subject of debate \cite{gyourko1989}.  Economists and other researchers have long documented the downsides of rent control policies \cite{minton1996}.  Those who argue against the policy cite unintended consequences such as reduced quality of housing, which can hurt property values, the decrease in the number of rental units available due to landlords converting apartments to condos, and the discouragement of new constructions \cite{Kenney2023}.  Only seven states currently have standing rent control policies and 37 states have bans against their implementation \cite{Kenney2023}.  On the other side of the debate, researchers argue that the second-generation rent control policies, known as ``rent stabilization" give local governments more freedoms and are more versatile \cite{pastor2018rent}.  Moreover, benefits from certain rent control policies have also been documented, such as reduced displacement and rent gauging.  A study of San Francisco's rent control policies has found that tenants are more likely to stay in the city as a direct consequence of the policy \cite{diamond2019effects}.  Another study focused on New Jersey showed that rent control policies helped prevent excessively high rent increases \cite{gilderbloom2007thirty}.  The question of whether the benefits of rent control outweigh the disadvantages is far from being settled.  The recent housing crisis that is occurring across the United States has reignited the rent control debate and states are beginning to consider removing the bans imposed on rent control.  For example, a new bill in Colorado 
would allow cities to pass their rent control laws \cite{Kenney2023}.  The study of the effects of rent control on displacement of vulnerable populations is thus timely and dynamical system models allow us to bring insight into this issue. 

\section{Model Specification}

\begin{figure*}[t]
    \centering
    \includegraphics[width = \textwidth]{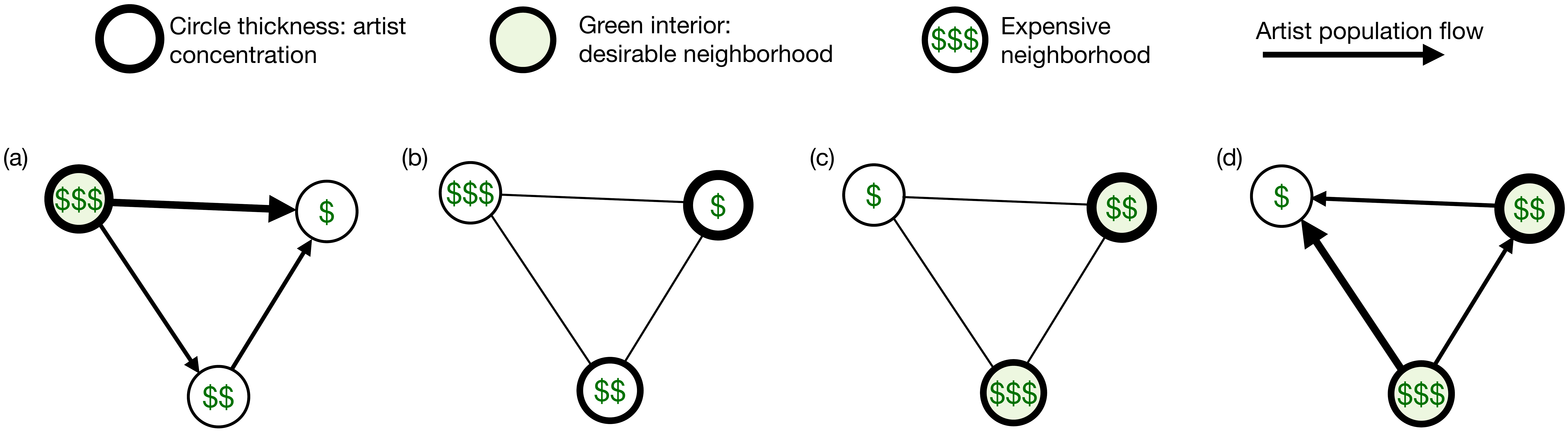}
    \caption{\justifying{(a) Artists transition from neighborhoods with higher prices to those with lower prices. (b) Consequently, these lower-priced neighborhoods experience an increase in their artist population. (c) The elevated artist population enhances the desirability and price of these neighborhoods. (d) The subsequent rise in prices drives artists to move to other, more affordable neighborhoods, thus repeating the cycle.}}
    \label{fig:cartoon2}
\end{figure*}

To model the process of gentrification in its simplest setting, we propose a model with only three variables in each neighborhood: the fraction of the artist population currently living in the neighborhood, $A$, the desirability of the neighborhood, $D$, and the average price of real estate in the neighborhood, $P$. Furthermore, for simplicity, in this paper, we do not consider heterogeneity in the neighborhoods (such as geographical location, amenities, or other intrinsic properties). According to the discussion above, our model will include the flow of the artist population towards lower-priced neighborhoods, the increase in the desirability of a neighborhood as its population of artists increases, and a subsequent increase in price as its desirability increases. As illustrated schematically in Fig.~\ref{fig:cartoon2}, our model makes the following assumptions: artists move from higher-priced to lower-priced neighborhoods [Fig.~\ref{fig:cartoon2}(a)]; as a result, these neighborhoods acquire a larger artist population [Fig.~\ref{fig:cartoon2}(b)] and later a higher desirability and price [Fig.~\ref{fig:cartoon2}(c)]; the resulting higher prices drive the artists towards other, lower-priced neighborhoods, repeating the cycle [Fig.~\ref{fig:cartoon2}(d)]. In our model, we assume that in the absence of artists, the desirability of a neighborhood decreases to a lower value. 

For modeling purposes we work at the scale of neighborhoods, each considered a distinct region. We assume there are $N\geq 2$ neighborhoods labeled $n = 1,2,\hdots, N.$ In this section, we introduce a model for the dynamics and interplay of the fraction of artists in neighborhood $n$, $A_n$, the desirability of the neighborhood, $D_n$,  and the average price of real estate in that neighborhood, $P_n$, in each of the $N$ neighborhoods.  We assume that an artist's decision to move from one neighborhood to another is determined by the difference in the average price of real estate in the two neighborhoods.  Specifically, they will move only to neighborhoods with lower prices.  For example, artists in a neighborhood $n$ with real estate price $P_n$ will move to neighborhood $k$ only if $P_k<P_n$ and, for simplicity, we assume that the flow rate is given by $h(P_n-P_k)A_n$, where
\begin{gather*}
h(x) = 
    \begin{cases}
    x, & \text{if } x\geq 0, \\
    0, & \text{otherwise}. 
    \end{cases}
\end{gather*}
According to our previous discussion, we assume that the influx of artists into a neighborhood leads to an increase in the desirability of the neighborhood. Subsequently, this higher desirability increases the average price of real estate.  These dynamics give rise to the following model
{\small
\begin{gather}\label{sys:model0}
    \left\{\begin{array}{ll}
     \tau_A\frac{dA_n}{dt} & = \sum\limits_{m=1}^N [h(P_m-P_n)A_m - h(P_n-P_m)A_n],\vspace{12pt}\\
    \tau_D\frac{dD_n}{dt} & = \sigma(A_n) - D_n,\vspace{12pt}\\
     \tau_P\frac{dP_n}{dt} & = D_n - P_n,
\end{array}\right.
\end{gather}}
for $t>0$ and with initial data $A_n(0)\geq 0, \ D_n(0)\geq 0,$  $P_n(0)\geq 0$, and $\sum_{n = 1}^N A_n(0) = 1$.
 The parameters $\tau_A$, $\tau_D,\ \tau_P \in \R^+$ are the characteristic timescales of artist migration, change in desirability, and change in real-estate prices of a region, respectively. The function $\sigma(\cdot)$ is a sigmoid function given by
\begin{equation}
\label{eq:sigma}
    \sigma(x) = \frac{1}{2}\left[1 + \tanh\left(\frac{x-z}{\epsilon}\right)\right],
\end{equation}
for positive real parameters $z$ and $\epsilon.$
The first equation in \eqref{sys:model0} describes the flow of artists between different neighborhoods. The term in the square brackets represents the flow of artists from neighborhood $m$ into neighborhood $n$ and the flow of artists from neighborhood $n$ into neighborhood $m$ according to the flow model proposed above.  The movement of artists can be understood as unidirectional diffusion, where diffusion occurs only towards neighborhoods with lower prices. The second equation describes how the desirability of neighborhood $n$, $D_n$, evolves given the fraction of artists $A_n$. When $A_n$ is larger than the threshold $z$ of the sigmoid function $\sigma$, the term $\sigma(A_n)$ becomes larger and causes $D_n$ to increase with timescale $\tau_D$. The parameter $\epsilon$ controls how sharply $\sigma$ increases with $A_n$. Finally, the third equation specifies that the real estate price of neighborhood $n$ relaxes to the current desirability with timescale $\tau_P$.  By rescaling time if necessary, from now on we will assume without loss of generality that $\tau_A = 1.$

System \eqref{sys:model0} is an extreme simplification of a complex phenomenon and makes many assumptions. In particular, it ignores any spatial organization and heterogeneity in the intrinsic desirability of neighborhoods, considers only one social group, and assumes very specific dynamics. Moreover, in many cases, once the average price of real estate has increased it will not decrease due to an outflow of artist populations. Nevertheless, our model is useful because it demonstrates, in the simplest setting, how similar but more realistic models could be postulated. It also shows that the interplay between real estate prices and the movement of population subgroups can lead to rich dynamic behavior. In Sec.~\ref{sec:discussionandconclusions} we discuss the limitations of the model and possible extensions in more detail. 

\section{Steady-state dynamics}\label{sec:observed behavior}
\begin{figure}
    \centering
    \begin{subfigure}[b]{0.4\textwidth}
    \includegraphics[width=\textwidth]{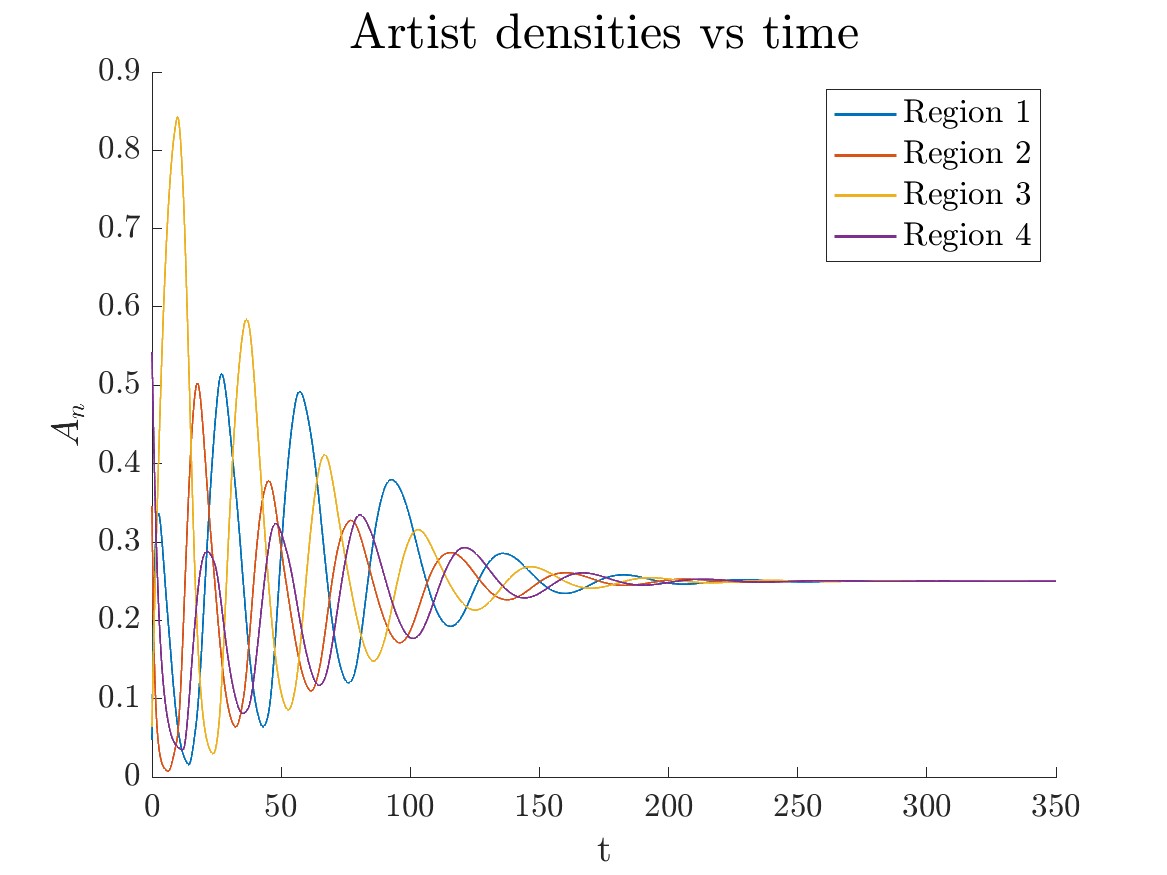}
    \caption{$N=4, \ \tau_D=5, \ \tau_P=5$}
    \label{fig:fixed point}
    \end{subfigure}

    \begin{subfigure}[b]{0.4\textwidth}
    \includegraphics[width=\textwidth]{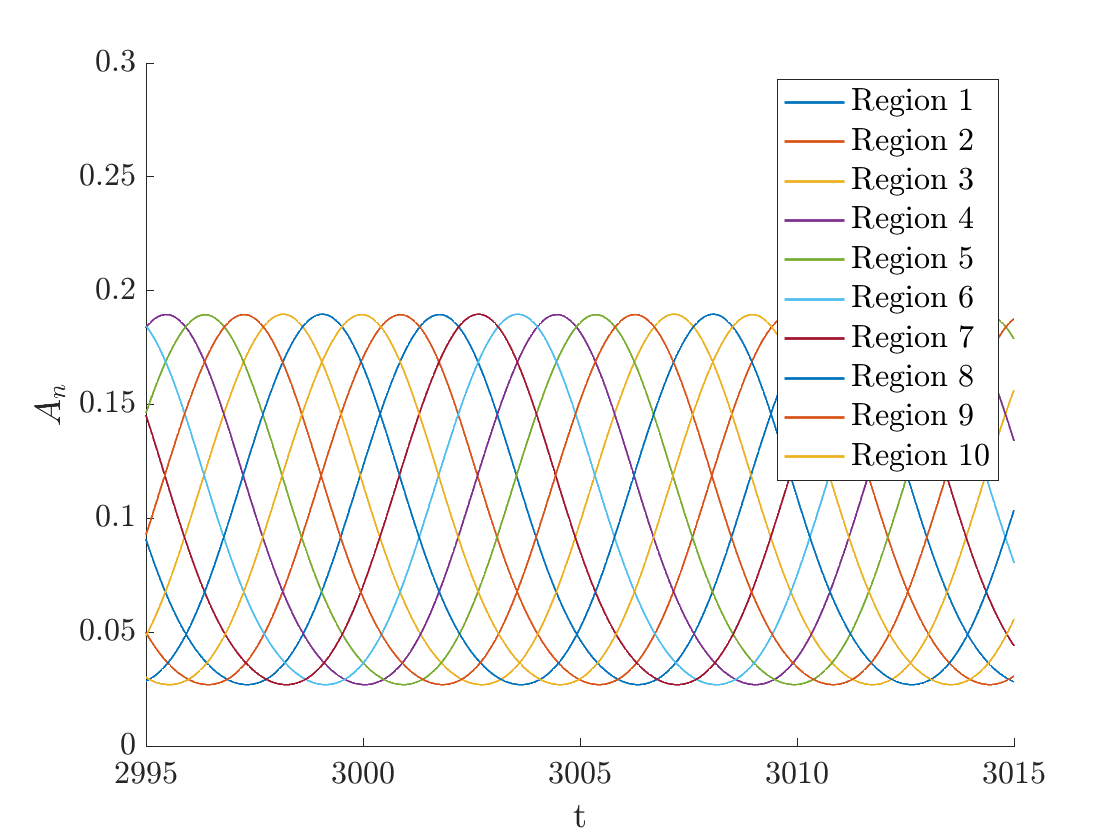}
    \caption{$N=10, \ \tau_D=2, \ \tau_P=2$}
    \label{fig:linearly spaced oscillations}
    \end{subfigure}

    \begin{subfigure}[b]{0.4\textwidth}
        \includegraphics[width=\textwidth]{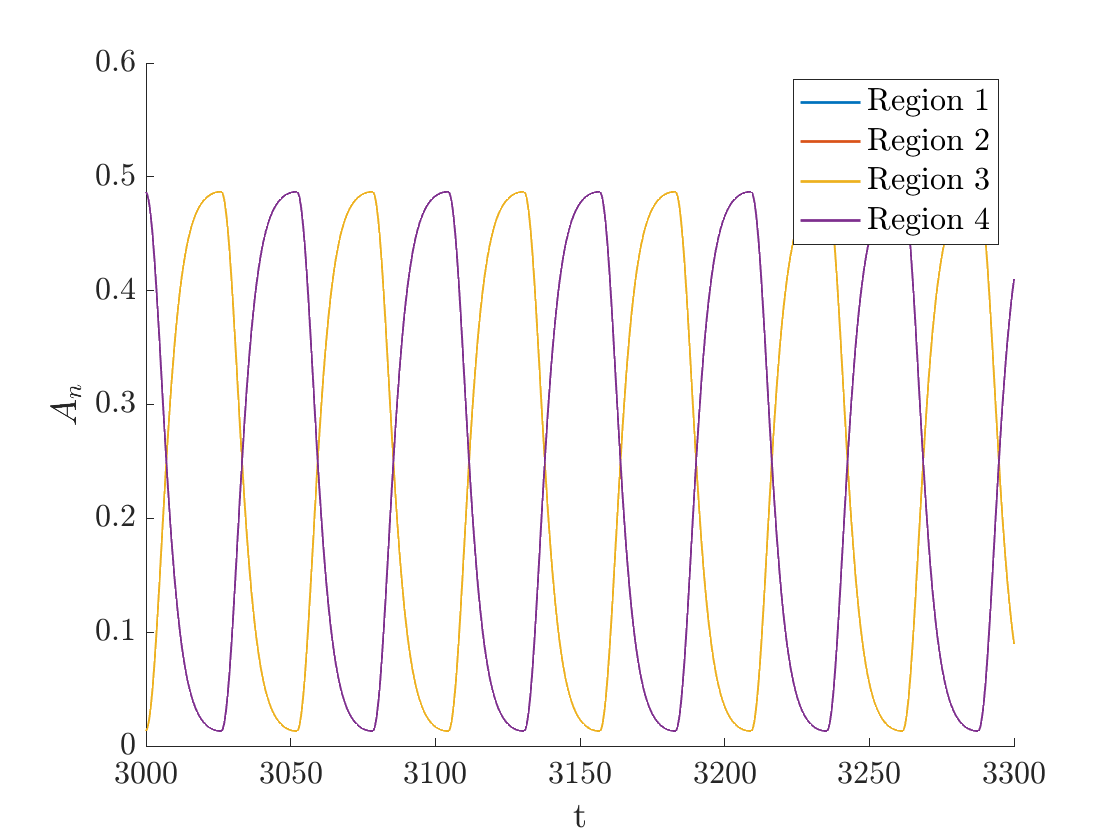}
      \caption{$N=4, \ \tau_D=15, \ \tau_P=15$} 
      \label{fig:grouped oscillations}
    \end{subfigure}

    \begin{subfigure}[b]{0.4\textwidth}
      \includegraphics[width=\textwidth]{disorgosc.png}
      \caption{$N=7, \ \tau_D=18, \ \tau_P=18$}
      \label{fig:disorganized oscillations}
    \end{subfigure}
    \caption{\justifying{Fraction of artists in region $n$ as a function of time for various parameter choices found by solving system \eqref{sys:model0}.}}
    \label{fig:observed behavior}
\end{figure}

Despite the simplicity of system \eqref{sys:model0}, the qualitative behavior of its solutions is rich. In Fig.~\ref{fig:observed behavior} we illustrate the four types of steady-state dynamics that we have numerically observed of the solutions to system \eqref{sys:model0}. 

In Fig.~\ref{fig:fixed point} we plot the fraction of artists $A_n$ for $n = 1,2,3,4$ for parameters $N = 4$, $\tau_D = 5$, $\tau_P = 5$. In this case, the system approaches an equilibrium $A_n=1/N.$ In Section \ref{sec:linear stability analysis} we will study the linear stability of this equilibrium. In Fig.~\ref{fig:linearly spaced oscillations}, which has parameters $N = 10, \ \tau_D = 2, \ \tau_P = 2,$ the population of artists in the regions alternate sequentially and appear to be lag-synchronized (see \cite{2002PhR...366....1B}): the population of artists satisfies
\begin{equation*}
    A_n(t) = A\left(t - \frac{Ti_n}{N}\right), \ \text{where}
\end{equation*}
$A$ is a function with period $T$ and $\{i_n\}_{n=1}^N$ is a permutation of $\{1,2,\hdots,N\}$. In this regime, artists become periodically displaced from the regions they move to, but the pattern of displacement is regular. In addition, each region has at some point in time the majority of the artist population. In Fig.~\ref{fig:grouped oscillations}, for parameters $N = 4, \ \tau_D = 15, \ \tau_P = 15,$ the regions split into two clusters $\{n_1,n_2\}, \ \{n_3,n_4\}$ such that 
\begin{equation*}
    A_{n_1}(t) = A_{n_2}(t) \ \text{and} \ A_{n_3}(t) = A_{n_4}(t), 
\end{equation*}
corresponding to cluster synchronization \cite{2002PhR...366....1B}. Finally, in Fig.~\ref{fig:disorganized oscillations}, for parameters $N = 7, \ \tau_D = 18, \ \tau_P = 18,$ the artist populations appear to oscillate erratically. The oscillations are quasiperiodic but with periods much longer than the relevant timescales $\tau_D, \ \tau_P$, and $\tau_A = 1$. Therefore, in the relevant timescales, the system's behavior appears to be highly disorganized. In Section~\ref{sec:rent-control} we will introduce a quantitative measure of “disorganization” and explore how the disorganization of the system depends on the system parameters. 

We note that the steady-state dynamics shown in Fig.~\ref{fig:observed behavior} occur, for the initial conditions we used, after a long transient time (see the horizontal axis scale). This time is so long that it is unrealistic even for gentrification process timescales. However, we argue that, under the assumptions of the model, knowledge of the nature of the long-term dynamics can be useful for assessing the stability of desired behavior and design interventions, such as encouraging a uniform distribution of the artist population [a stable situation for the parameters of Fig.~\ref{fig:fixed point}], or promoting a regular alternation of desirability between neighborhoods [stable for the parameters of Fig.~\ref{fig:linearly spaced oscillations}]. 

\section{Linear Stability Analysis of the Homogeneous Equilibrium Solution}\label{sec:linear stability analysis}
In this section, we find the equilibria of system \eqref{sys:model0} and study their linear stability. 

\subsection{Linear stability of the spatially homogeneous equilibrium solution}
Since $A_n$ denotes the fraction of artists in neighborhood $n$, then $\sum_{n = 1}^N A_n = 1.$ By symmetry, one expects that an equilibrium solution satisfies $A_n = 1/N$. One can check directly that $A_n = A^* = 1/N, \ D_n = D^* = \sigma(1/N),$ and $P_n = P^* = \sigma(1/N)$ for all $n$ is a solution of system \eqref{sys:model0}. In addition, one can show that this is the only equilibrium solution (see Appendix \ref{Appendix A}). 

In what follows we consider the state of the system to be described by the state vector 
\begin{equation*}
    {\bf x} = \left[A_1,A_2,\hdots,A_N,D_1,D_2,\hdots,D_N,P_1,P_2,\hdots,P_N\right]^\mathrm{T}
\end{equation*}
and perform a linear stability analysis of the spatially homogeneous solution, given by the vector
\begin{equation} \label{eq:equilibrium state vec}
{\bf x^*} = 
    \begin{bmatrix}
        {\bf A^*} \\ 
        {\bf D^*} \\
        {\bf P^*}
    \end{bmatrix},
\end{equation}
with ${\bf A^*} = [A^*,A^*,\hdots,A^*]^\mathrm{T}$,  ${\bf D^*} = [D^*,D^*,\hdots,D^*]^\mathrm{T}$, and  ${\bf P^*} = [P^*,P^*,\hdots,P^*]^\mathrm{T} \in \R^N.$ The linear stability of this solution is determined by the eigenvalues of the Jacobian associated with system \eqref{sys:model0} evaluated at ${\bf x^*}$, given by the block matrix
\begin{align} \label{eq:M matrix}
    M = 
    \begin{bmatrix}
      0 & 0 & M_{13} \\
      M_{21} & M_{22} & 0 \\
      0 & M_{32} & M_{33}
    \end{bmatrix},
\end{align}
where $M_{13}$ is an $N$-dimensional block given by
\begin{align*}
M_{13} = 
\begin{bmatrix}
(1-N)A^* & A^* & \hdots & A^*  \\[5pt]
A^*  & (1-N)A^* & \hdots & A^*  \\[5pt]
\vdots & \vdots & \ddots & \vdots \\[5pt]
A^*  & A^*  & \hdots & (1-N)A^*
\end{bmatrix},
\end{align*}
and
\begin{align*}
    \begin{cases}
    M_{21} = \frac{\sigma '(A^*)}{\tau_D} \mathbb{I}, \\[5pt]
    M_{22} = -\frac{1}{\tau_D} \mathbb{I} ,\\[5pt]
    M_{32} = \frac{1}{\tau_P} \mathbb{I}, \\[5pt]
    M_{33} = -\frac{1}{\tau_P} \mathbb{I},
    \end{cases}
\end{align*}
where $\mathbb{I}$ is the $N\times N$ identity matrix. 
Three eigenvalues of $M$ are
    $$
 \lambda_1 = 0,\;\lambda_2 = -\frac{1}{\tau_D},\;\text{and}\; \lambda_3 = -\frac{1}{\tau_P},
$$
with algebraic multiplicity one. The eigenvalue $\lambda_1 = 0$ is associated with a uniform perturbation to the artist populations across all regions, i.e., such that $\delta A_n =\delta A$ is independent of $n.$ Such perturbations are not permitted since they would violate the condition [supported by the fact that $\sum_n A_n$ is conserved by system \eqref{sys:model0}] that $\sum_{n=1}^NA_n = 1$. Therefore, since $\lambda_2$ and $\lambda_3$ are negative, the linear stability of the equilibrium is determined by the remaining eigenvalues. In Appendix \ref{Appendix B} we show that the remaining eigenvalues of $M$ are solutions to the equation 
\begin{equation*}
    \left[\frac{\sigma'(A^*)}{ \tau_D \tau_P }+\lambda \left(\frac{1}{\tau_D} + \lambda \right) \left(\frac{1}{\tau_P} + \lambda \right) \right]^{N-1} = 0,
\end{equation*}
from which the linear stability condition
\begin{equation}
\label{eq:stability condition}
    \sigma'(A^*) < \frac{1}{\tau_D} + \frac{1}{\tau_P}
\end{equation}
can be derived. Equation \eqref{eq:stability condition} shows that the homogeneous equilibrium becomes unstable if either the timescales of desirability and price changes are too large, or if the sigmoid determining how the artist population changes the desirability of a neighborhood is sufficiently steep.

\subsection{Numerical verification of the linear stability analysis}

 Now, we numerically verify the linear stability condition given in Eq.~\eqref{eq:stability condition}. For simplicity, for the remainder of the paper, we fix the parameters of the function $\sigma$ to be $z = 0.01$ and $\epsilon = 0.1$ and explore the behavior that system \eqref{sys:model0} exhibits as a function of $N$, $\tau_D$, and $\tau_P$. Additionally, we assume $\tau_D, \tau_P \in (0,20]$, as slower response rates to artist population densities would not correspond to rapid gentrification.  To verify condition \eqref{eq:stability condition}, we calculated numerically the growth rate of perturbations from the equilibrium solution \eqref{eq:equilibrium state vec}. For a given set of parameters $N, \ \tau_D, \ \tau_P,$ we added a small random perturbation ${\bf \delta x}(0)$ to $\bfx^*$ and let it evolve using Eqs.~\eqref{sys:model0}. The growth rate was estimated as $\lambda_R \approx \ln\left(\|{\bf \delta x}(t_w)\|/\|{\bf \delta x}(0)\|\right)/t_w$, where $t_w$ is chosen large enough that perturbations have time to grow or decay, but short enough that their size does not saturate.

\begin{figure}[t]
    \centering
    \includegraphics[width = 0.4\textwidth]{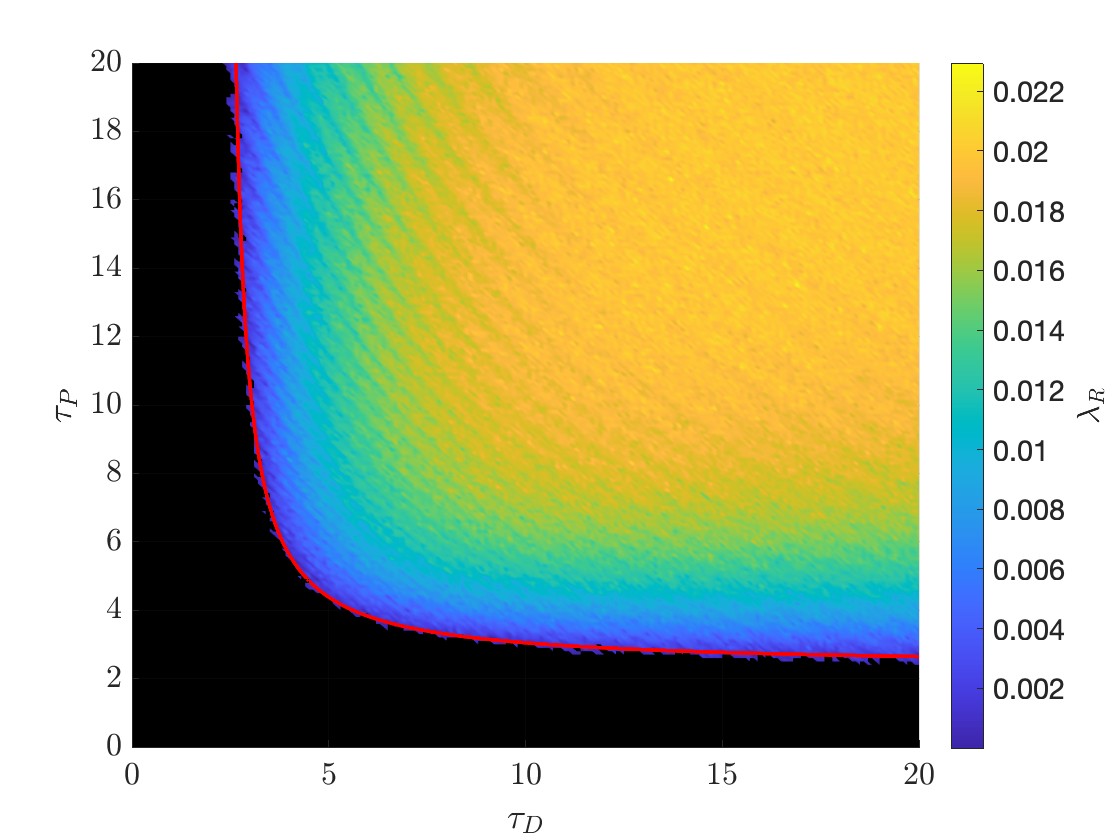}
    \caption{\justifying{Growth rate of perturbations $\lambda_R$ from the equilibrium solution $A_n = 1/N$ calculated numerically as a function of $\tau_D$ and $\tau_P$ for the $N = 5$ case.}}
    \label{fig:perturbation heat map}
\end{figure}

  Figure \ref{fig:perturbation heat map} shows the growth rate $\lambda_R$ for the $N = 5$ case. The black region of the figure indicates a parameter regime under which the perturbed system returned to the spatially homogeneous solution (i.e., $\lambda_R<0$), and the red line indicates the linear stability condition given in \eqref{eq:stability condition}. As expected, the linear stability condition \eqref{eq:stability condition} accurately predicts the growth or decay of perturbations to the equilibrium solution \eqref{eq:equilibrium state vec}.

\begin{figure}[b]
    \centering
    \includegraphics[width=0.4\textwidth]{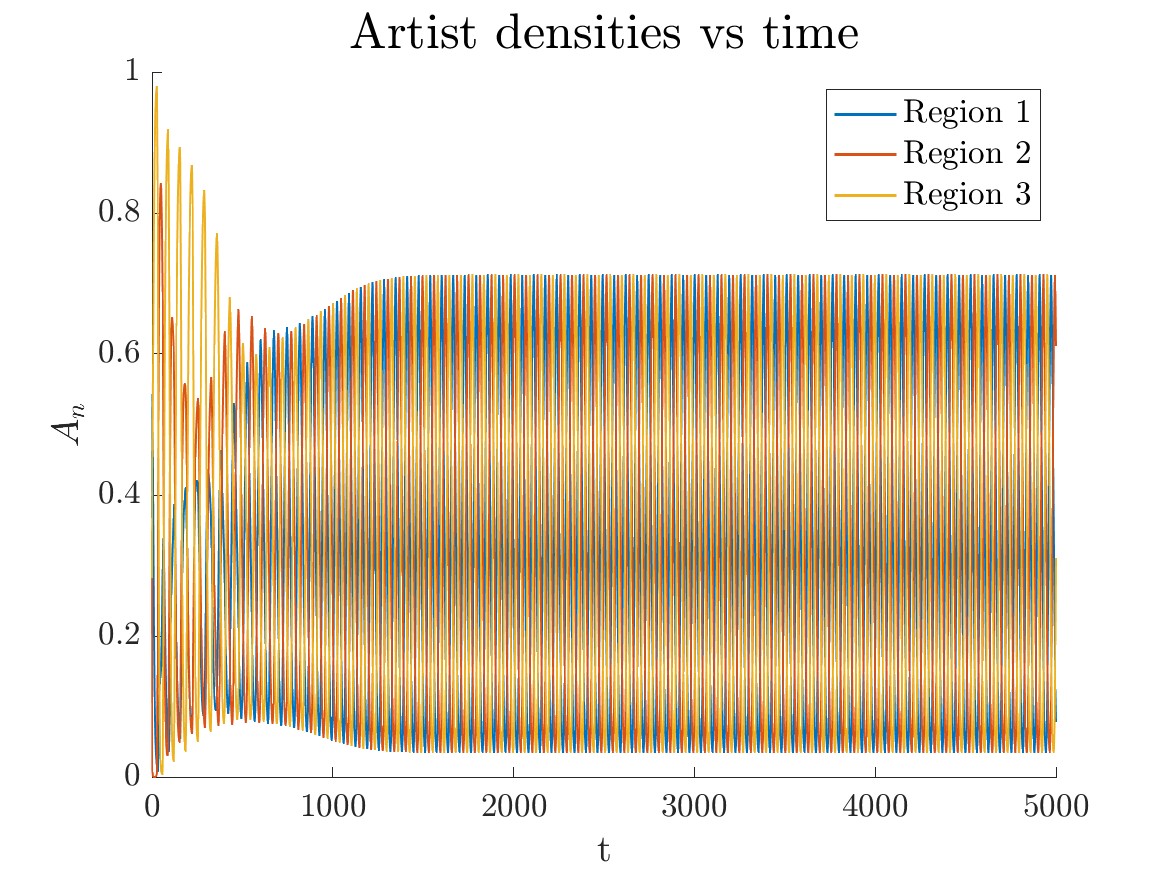}
    \caption{\justifying{Artist populations $A_n$ versus time $t$ for $N=3, \ \tau_D=15, $ and $\tau_P = 15$ for parameter values where the homogeneous equilibrium solution is linearly stable. }}
    \label{fig:no convergence to the fixed point}
\end{figure}

Condition \eqref{eq:stability condition} guarantees the linear stability of the equilibrium solution given by \eqref{eq:equilibrium state vec}. However, it does not guarantee global stability. Therefore, it is possible that solutions to system \eqref{sys:model0} do not converge to the spatially homogeneous equilibrium solution under the linearly stable parameter regime. A case illustrating this is shown in Fig.~\ref{fig:no convergence to the fixed point}, which shows the artist populations $A_n$ versus time $t$ for $N = 3, \ \tau_D = 15,$ and $\tau_P = 15$.  For these parameters, condition \eqref{eq:stability condition} is satisfied but oscillations of the form shown in Fig.~\ref{fig:linearly spaced oscillations} persist on a long timescale. 

\begin{figure}[t]
    \centering
    \includegraphics[width = 0.4\textwidth]{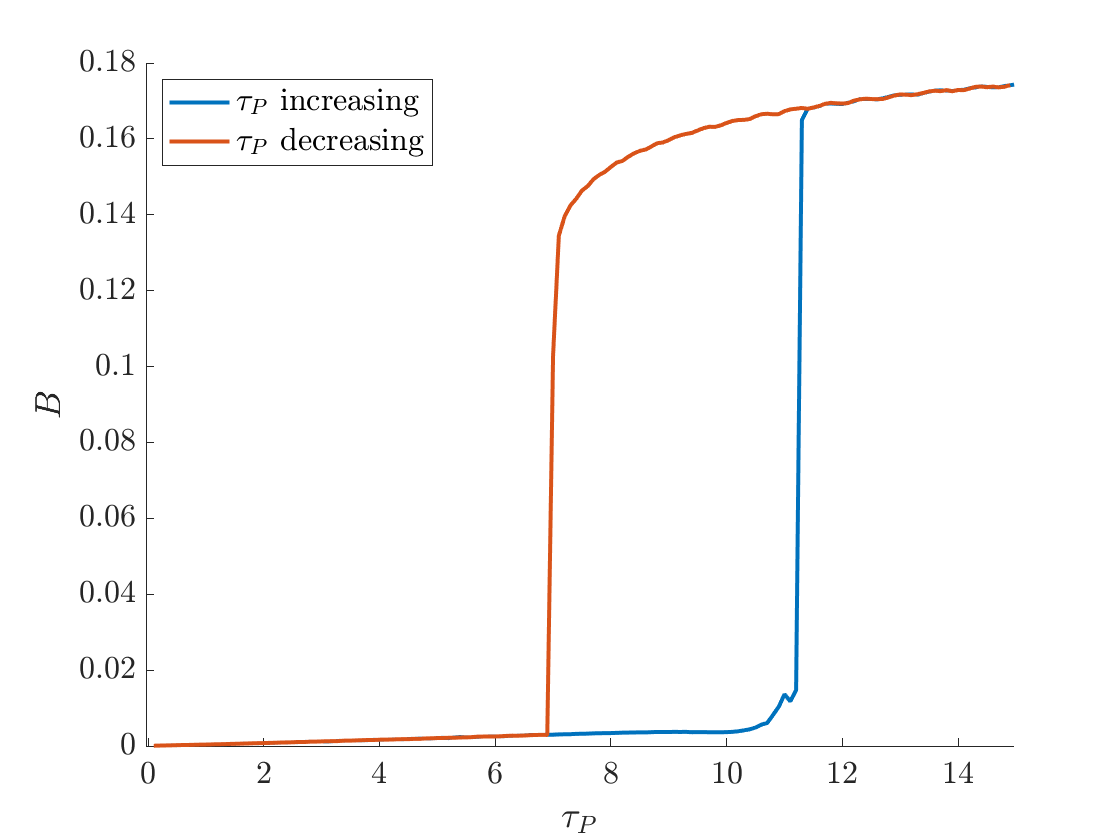}
    \caption{\justifying{Abrupt transition of solution from the equilibrium solution to lag-synchronized oscillations at $\tau_P \approx 11.1$ when increasing $\tau_P$ (blue curve) and vice-versa at $\tau_P \approx 7.1$ when decreasing $\tau_P$ (orange curve) for $\tau_D = 15$ and $N = 4$.}}
    \label{fig:bistability}
\end{figure}

Systems with $N = 3$ and 4 admit bistability, in which equilibrium and lag-synchronized oscillatory solutions are simultaneously stable. We illustrate this in Fig.~\ref{fig:bistability} for the $N = 4$ and $\tau_D = 15$ case. Starting with initial conditions close to the homogeneous equilibrium solution, we slowly increased $\tau_P$ from $0.1$ to $15$ in steps of $0.1$. To characterize the proximity of the state to the equilibrium solution over time, we calculate the distance $B$ of the artist populations $A_n$ to the equilibrium $1/N$ averaged over time and over regions,  
\begin{equation}
    B = \frac{1}{N} \sum_{n = 1}^N \left\langle \left|A_n - \frac{1}{N}\right|\right\rangle_t, 
\end{equation}
where $\langle \cdot \rangle_t$ indicates a time average. 
For each set of parameters $(\tau_D,\tau_P)$ in the aforementioned range, system \eqref{sys:model0} is numerically advanced 15,000 time units, and the statistic $B$ is computed after discarding transients. As shown in Fig.~\ref{fig:bistability}, the system remains close to the equilibrium solution up to approximately $\tau_P \approx 11.1,$ at which point this solution becomes unstable and the system transitions abruptly to lag-synchronized oscillations. Subsequently decreasing $\tau_P$, we find that the lag-synchronized oscillations persist down to approximately $\tau_P \approx 7.1$, indicating a bistable regime in which both lag-synchronized oscillations and the uniform equilibrium solution are stable. 

\subsection{Organization of steady-state oscillatory solutions}

In Sec.~\ref{sec:observed behavior} we described the four types of steady-state behavior we have observed: (i) a homogeneous equilibrium where $A_n(t) = 1/N$ (Fig.~\ref{fig:fixed point}), (ii) lag-synchronized behavior where the artist populations in different regions oscillate with uniform phase lags (Fig.~\ref{fig:linearly spaced oscillations}), (iii) cluster-synchronized behavior where the populations split into synchronized groups that undergo identical oscillations, and (iv) quasiperiodic behavior which, on the timescales of interest, appears disorganized. 

Our goal in this section is to numerically determine the parameter regimes leading to the different behaviors of the solutions to system \eqref{sys:model0}. For simplicity, we will distinguish three cases: an equilibrium [case (i)], organized oscillations [cases (ii) and (iii)], and disorganized oscillations [case (iv)]. We note that organized oscillations are characterized by the fact that all local maxima are equal, both across regions and over time. To quantify this, we introduce a statistic $\chi$ defined as
\begin{equation}\label{eq:chi def}
    \chi  = \frac{2}{N}\sum_{n=1}^N\frac{\sigma_n}{m_n},
\end{equation}
where $\sigma_n$ is the standard deviation of the local maxima in the time series of the artist population in the region $n,$ $A_n(t)$, and $m_n$ is the mean of $A_n(t)$. Values $\chi\approx 1$ indicate that the variation in the amplitude of artist population oscillations is comparable to their mean, while $\chi \ll 1$ approximately indicates either an equilibrium solution or organized oscillations, which we identify separately (see Appendix~\ref{Appendix C}). We note that, in principle, $\chi\approx 1$ could also indicate other relatively organized behavior such as period-2 oscillations, but we have not observed these. 
Figure~\ref{fig:observed behavior} lends some intuition to $\chi$, as Figs.~\ref{fig:linearly spaced oscillations}, \ref{fig:grouped oscillations}, and \ref{fig:disorganized oscillations} respectively yield $\chi$ values of approximately $ 4.8\times10^{-6}$, $ 5.1\times10^{-8}$, and $ 0.43$. Equilibria are identified numerically by the absence of distinct local maxima or by the condition std$\left\{A_n(t)\right\}< 10^{-3}$.

\begin{figure}
    \centering
    \begin{subfigure}[b]{0.4\textwidth}
    \includegraphics[width=\textwidth]{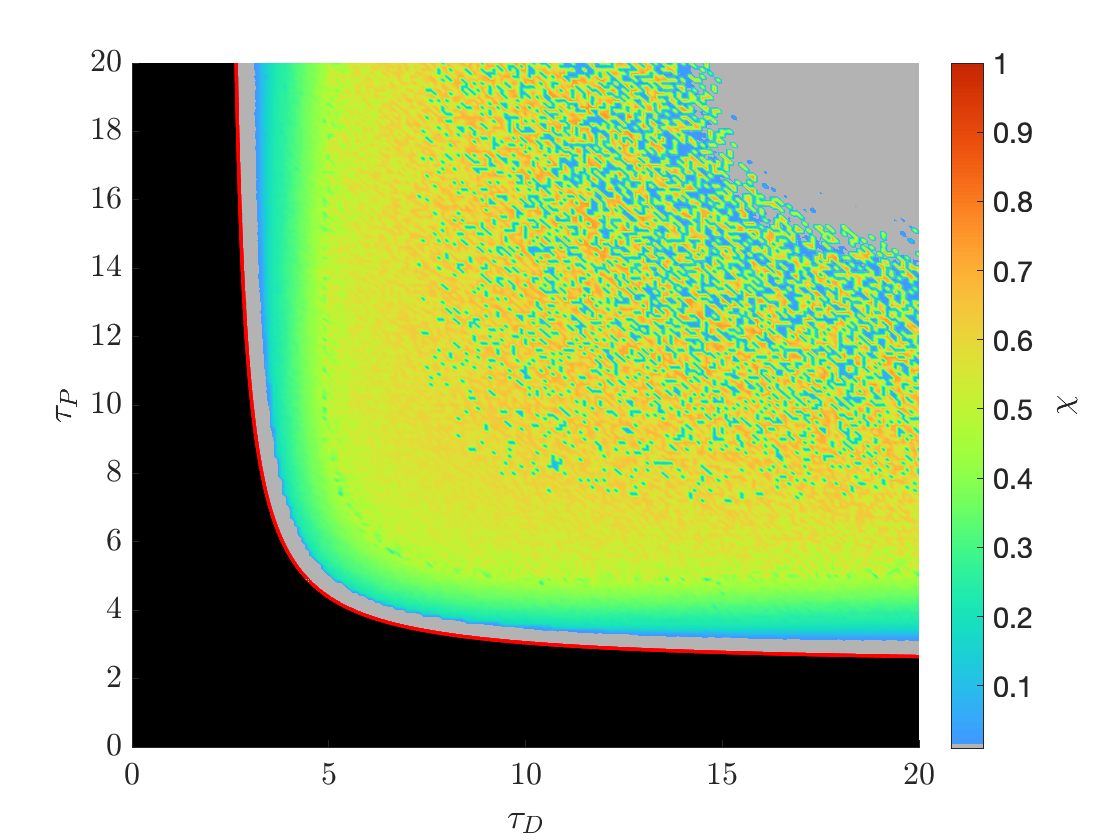}
    \caption{$N=5$}
    \end{subfigure}

   \begin{subfigure}[b]{0.4\textwidth}
    \includegraphics[width=\textwidth]{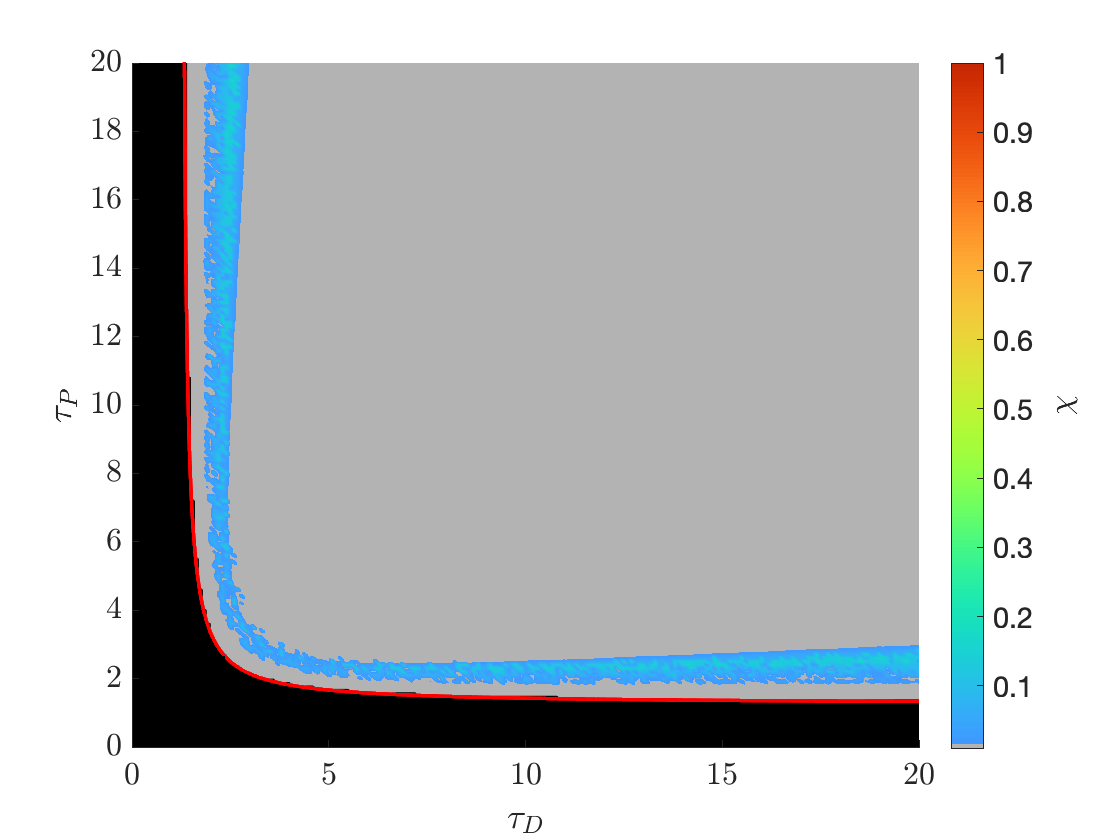}
    \caption{$N=6$}
    \end{subfigure}

   \begin{subfigure}[b]{0.4\textwidth}
    \begin{tikzpicture}
    \node[anchor=south west,inner sep=0] (image) at (0,0) {\includegraphics[width=\textwidth]{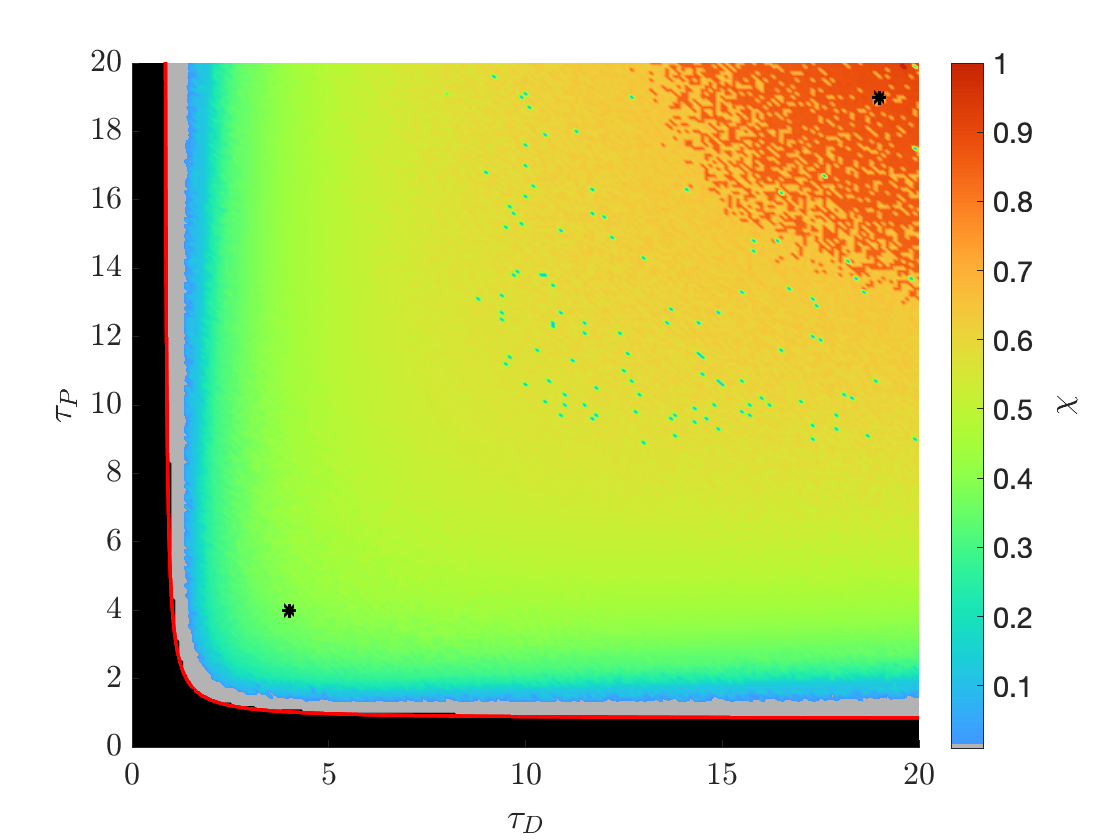}};
    \begin{scope}[x={(image.south east)},y={(image.north west)}]
\draw[->] (0.31,0.4)--(0.27,0.3); 
\node at (0.3,0.45) {Fig.~\ref{fig:neq7 chi pt1}};
\draw[->] (0.65,0.7)--(0.77,0.86); 
\node at (0.65,0.65) {Fig.~\ref{fig:neq7 chi pt2}};
    \end{scope}
\end{tikzpicture}
 \caption{$N=7$}
 \label{fig:neq7 chi pts}
    \end{subfigure}

    \begin{subfigure}[b]{0.4\textwidth}
    \includegraphics[width=\textwidth]{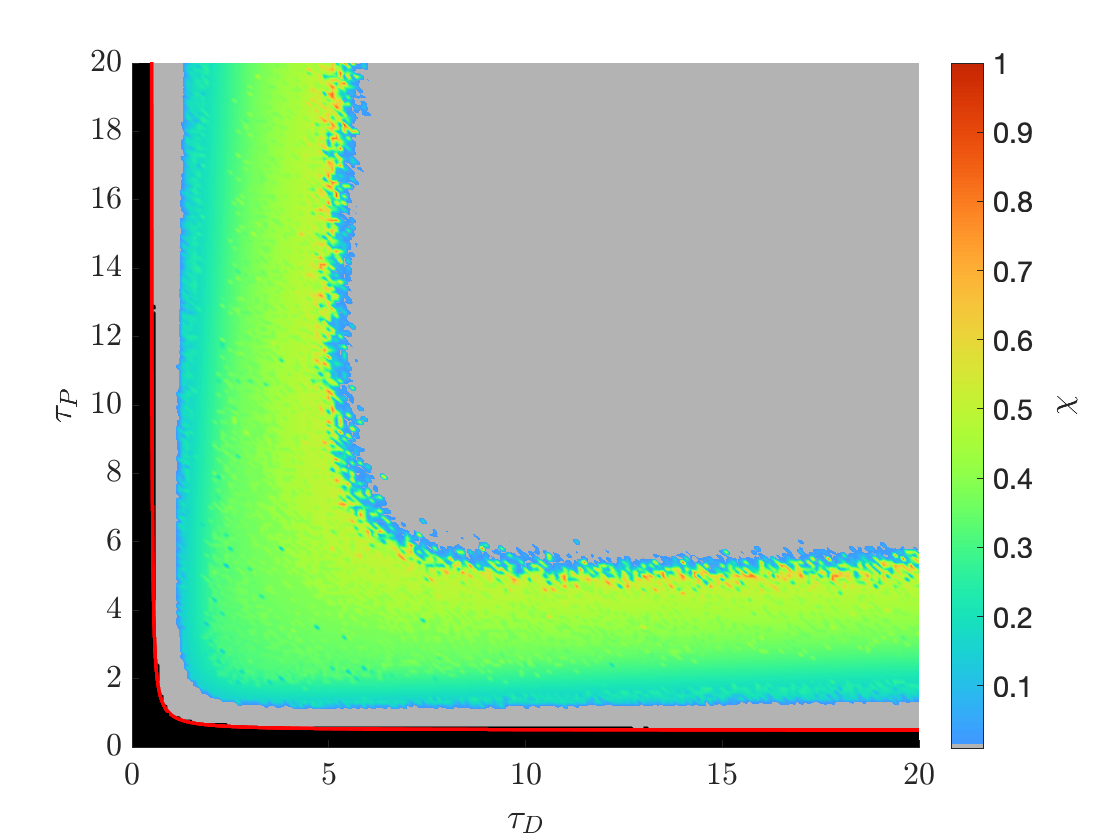}
    \caption{$N=9$}
    \label{fig:chi map, N=9}
    \end{subfigure}
    \caption{\justifying{The value of $\chi$ in the $N=5,6,7,$ and 9 cases.}}
    \label{fig:chi maps}
\end{figure}
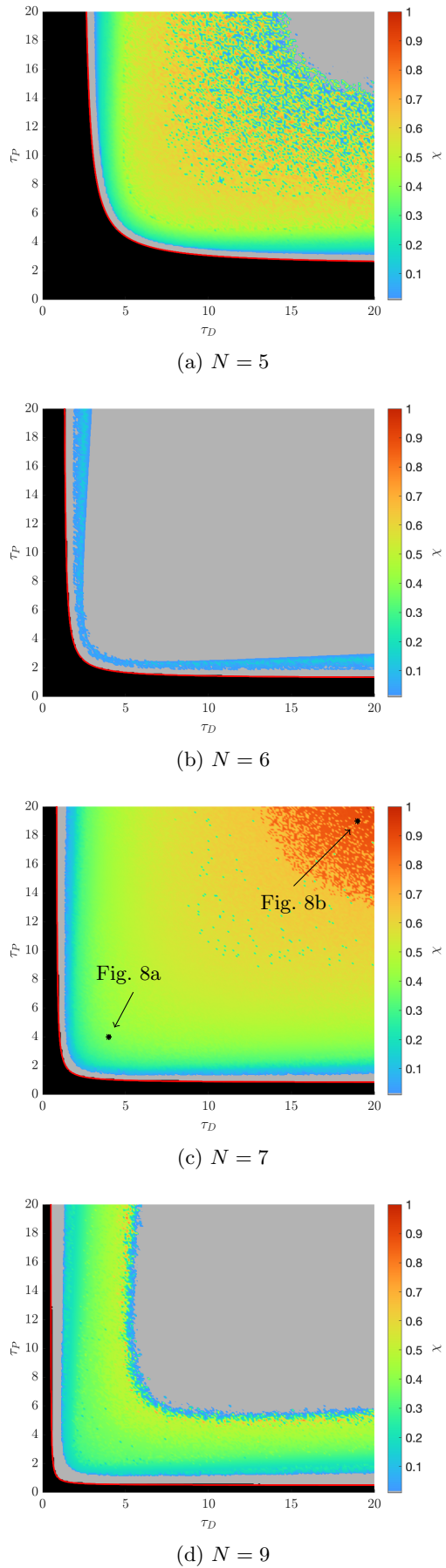

With the quantity $\chi$ we can visualize the degree of organization of steady-state oscillatory solutions in $\tau_D$-$\tau_P$ space, as shown in Fig.~\ref{fig:chi maps} for the $N = 5,6,7$ and 9 cases. As discussed above, we distinguish between equilibrium solutions [case (i)], organized oscillations [cases (ii) and (iii)], and disorganized oscillations. Black regions indicate an equilibrium solution, identified numerically as described above. Grey regions correspond to oscillations with a value of $\chi$ less than 0.01, which we identify as organized oscillations. Regions colored according to the color bar correspond to oscillations with a value of $\chi$ larger than 0.01, i.e., disorganized oscillations. As discussed previously, we find that organized oscillations typically have values of $\chi$ less than $10^{-3}$, so our choice of a 0.01 threshold effectively separates disorganized oscillations from the other observed dynamical regimes.  The red line illustrates the linear stability condition given by Eq.~\eqref{eq:stability condition}, as before. 

\begin{figure}
    \centering
    \begin{subfigure}[b]{0.4\textwidth}
    \includegraphics[width=\textwidth]{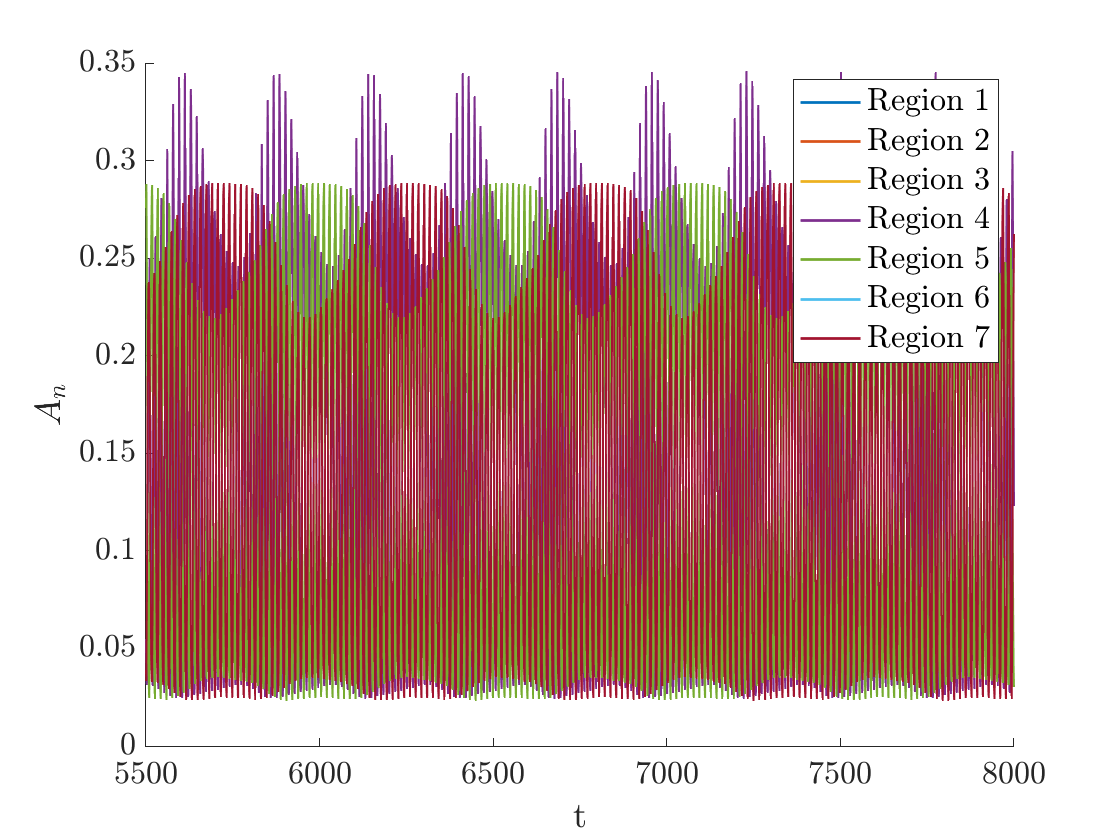}
    \caption{$N = 7, \ (\tau_D,\tau_P) = (4,4)$}
    \label{fig:neq7 chi pt1}
    \end{subfigure}

   \begin{subfigure}[b]{0.4\textwidth}
    \includegraphics[width=\textwidth]{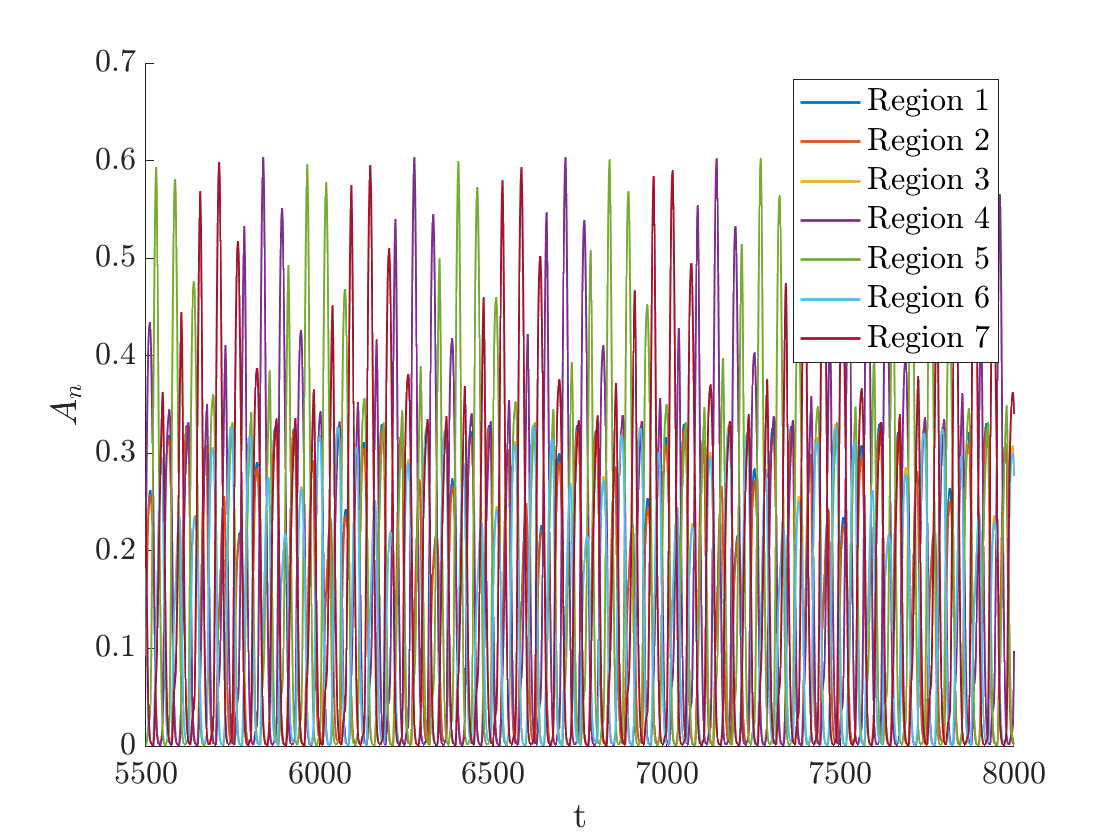}
    \caption{$N = 7, \ (\tau_D,\tau_P) = (19,19)$}
    \label{fig:neq7 chi pt2}
    \end{subfigure}
    \caption{\justifying{Time series corresponding to the two sets of parameters indicated in Fig.~\ref{fig:neq7 chi pts}.}}
    \label{fig:macroscopic organization of time series}
\end{figure}

Interestingly, the geometry of the  $\chi$ heat maps varies widely as a function of $N$. In particular, $\chi$ does not increase monotonically with Euclidean distance from the origin in $\tau_D$-$\tau_P$ space and instead admits maxima in a highly irregular fashion. Red parameter regimes ($\chi \approx 1$) typically give rise to strongly disorganized time series, while smaller values of $\chi$ indicate that oscillations are macroscopically organized. This is illustrated in Fig.~\ref{fig:macroscopic organization of time series}, which shows the time series for $N=7$ and $(\tau_D,\tau_P)$ equal to $(4,4)$ (top) and $(19,19)$ (bottom), with $\chi$ values of approximately $0.381$ and $0.886$, respectively. These parameter values are shown as black dots in Fig.~\ref{fig:neq7 chi pts}. 

Whether system \eqref{sys:model0} gives rise to chaotic time series is of particular interest, as this would indicate the continual and unpredictable displacement of the artist population. The $\chi$ heat maps presented above help constrain our search for chaotic solutions, as we do not expect that the parameter regions giving rise to small values of $\chi$ will support chaotic time series. Consequently, we subject the time series produced by system \eqref{sys:model0} for $N = 7, \ (\tau_D,\tau_P) \in [16,20] \times [16,20]$ to the 0-1 test for chaos \cite{Skokos2016}, as $\chi$ is maximized under this parameter regime. Solutions were generated for five sets of random initial conditions on a uniform grid mesh with spacing 0.1 for $(\tau_D,\tau_P) \in [16,20]\times[16,20]$, and the time series of artists across all regions were subjected to the 0-1 test. While the time series appears to be highly disorganized, we could not detect any chaos based on the 0-1 test. We conjecture that system \eqref{sys:model0} does not produce chaotic time series with significant amplitude variance for $z = 0.01, \ \epsilon = 0.1, \ N \in \{2,3\hdots,10\}, \ (\tau_D,\tau_P) \in (0,20]\times(0,20].$

\section{Implementing a rent-control strategy}\label{sec:rent-control}
The discussion in Section \ref{sec:rc} motivates the need to further study the benefits and drawbacks of rent control policies.    
Under the assumption that policymakers and city planners aim to maximize stability and minimize displacement, we analyze the effect that rent regulation policies have on the stability of neighborhood characteristics and the displacement of disadvantaged populations. 
In this Section, we consider a simple rent control strategy, coming from the first generation of rent control strategies implemented in the United States and in Europe, which is a rent cap that is uniformly applied. 
Mathematically, this is translated to a simple change of the dynamics of system \eqref{sys:model0}. In particular, we introduce a new variable $\hat P_n$ that represents the capped real estate (rent) price. As a first approximation, we assume that the market price, $P_n,$ continues to be subject to the dynamics modeled in system \eqref{sys:model0}. While in reality real estate prices are affected by rent control policies, here we explore only the effect of neighborhood desirability on prices. In neighborhoods with rent control policies in place, the actual cost (the one that enters the first equation and determines artists' decisions to move) is given by $\hat P_n= \min(P_n, P_n^{\text{max}}),$ where $P_n^{\text{max}}$ is the maximum price allowed in the neighborhood $n.$ To begin our investigation of these systems, we assume a single ceiling price, $P^{\text{ceil}}$, and assume the same rent control policy is in place for every neighborhood. Thus, the new dynamical system implementing this basic form of rent control is given by

{\small
\begin{gather}\label{sys:model1}
    \left\{\begin{array}{ll}
     \tau_A\frac{dA_n}{dt} & = \sum\limits_{m=1}^N [h(\hat{P}_m-\hat{P}_n)A_m - h(\hat{P}_n-\hat{P}_m)A_n],\vspace{12pt}\\
    \tau_D\frac{dD_n}{dt} & = \sigma(A_n) - D_n,\vspace{12pt}\\
     \tau_P\frac{dP_n}{dt} & = D_n - P_n,
\end{array}\right.
\end{gather}}

\noindent where $\hat{P}_n = \min(P_n,P^{\text{ceil}})$. 

\subsection{Equilibrium solutions and bounds}

System \eqref{sys:model1} gives rise to infinitely many equilibrium solutions. By the same reasoning presented in Appendix \ref{Appendix A}, an equilibrium solution to system \eqref{sys:model1} must satisfy $\hat{P}_1 = \hat{P}_2 = \hdots = \hat{P}_N.$ Therefore, the original spatially homogeneous solution given by Eq.~\eqref{eq:equilibrium state vec} continues to be an equilibrium solution for this system, but it is no longer a unique equilibrium, as any set of values $P_1, P_2, \hdots, P_N \geq P^{\text{ceil}}$ will give  $\hat{P}_1 = \hat{P}_2 = \hdots = \hat{P}_N$, disincentivizing artists to move. Consequently, since $\sigma(\cdot)$ is monotonically increasing, an equilibrium of system \eqref{sys:model1} is attained whenever 
\begin{align}
\label{eq:rent control equilibrium}
   \mathbf{A^*} = 
    \begin{pmatrix}
    A_1\\
    A_2\\
    \vdots \\
    A_N
    \end{pmatrix}, \ 
    \mathbf{D^*} = \sigma(\mathbf{A^*}), \
    \mathbf{P^*} = \mathbf{D^*}, 
\end{align}
for any $A_1, A_2, \hdots, A_N \geq \sigma^{-1}(\Pceil)$ with $\sum_{n=1}^N A_n=1.$ Values of $\mathbf{A^*}$, $\mathbf{D^*}$, and $\mathbf{P^*}$ satisfying (\ref{eq:rent control equilibrium}) form a continuum region $S$ of equilibria of the system with rent control. As expected for a continuum of equilibria, it can be shown that equilibria in the interior of $S$ are marginally stable (see Appendix \ref{Appendix E}).

As an example, Fig.~\ref{fig:Pequil values} shows $P_n(t)$ versus time for $N = 7$ and $\tau_D = \tau_P = 5$, $P^{\text{ceil}} = 0.8$. The red dashed line shows the value of $P^{\text{ceil}}$. After an initial transient, the prices stabilize at equilibrium values satisfying (\ref{eq:rent control equilibrium}). 
\begin{figure}[t]
    \centering
    \includegraphics[width = 0.4\textwidth]{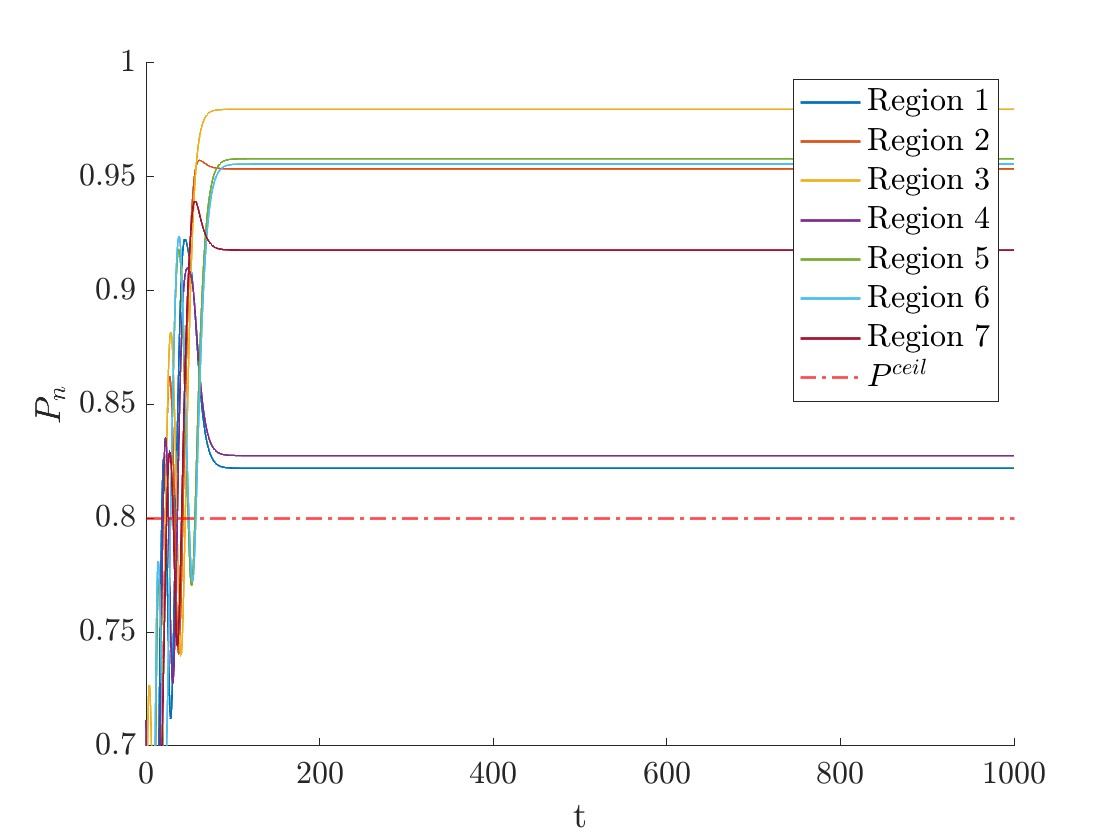}
    \caption{Approach to equilibrium prices in the $N = 7, \ \tau_D = 5, \ \tau_P = 5, \ \Pceil = 0.8$ case.}
    \label{fig:Pequil values}
\end{figure}
\begin{figure}[b]
    \centering
    \includegraphics[width = 0.4\textwidth]{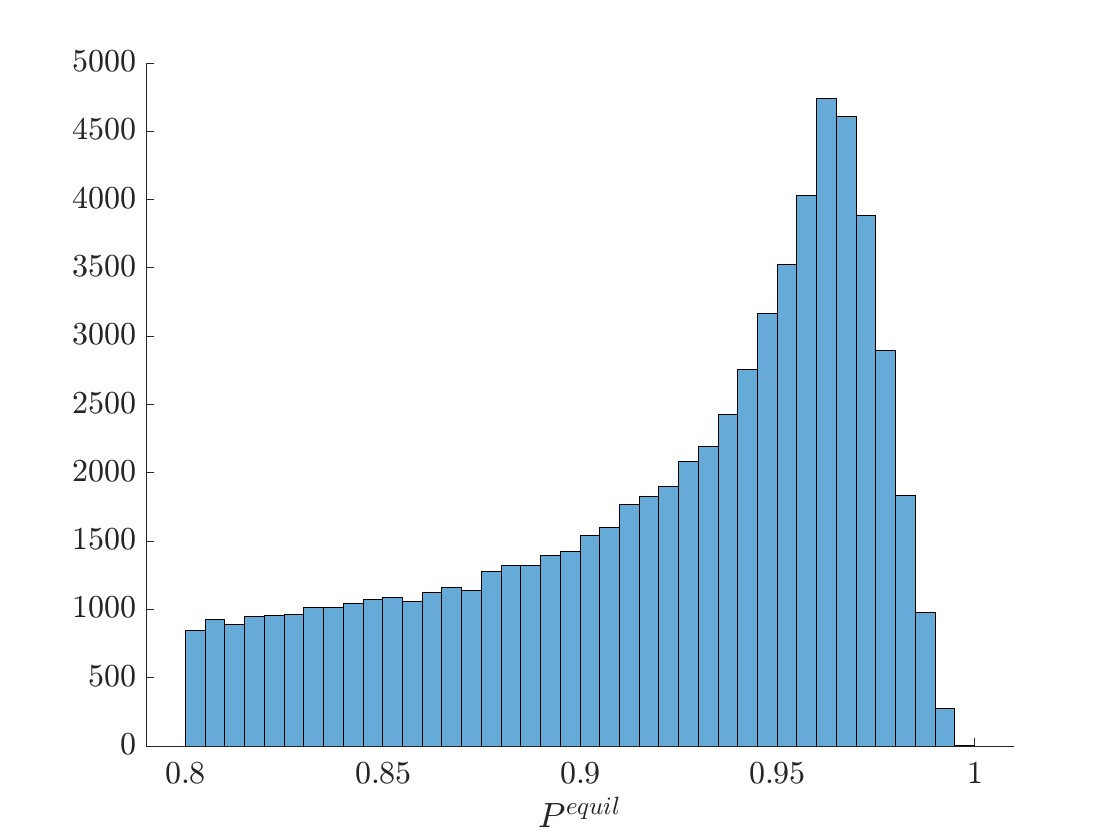}
    \caption{Histogram of equilibrium price values for the $N = 7, \ \tau_D = 5, \ \tau_P = 5, \ \Pceil = 0.8$ case.}
    \label{fig:Pequil histogram}
\end{figure}
\begin{figure}
\centering
    \begin{subfigure}[b]{0.4\textwidth}
    \includegraphics[width=\textwidth]{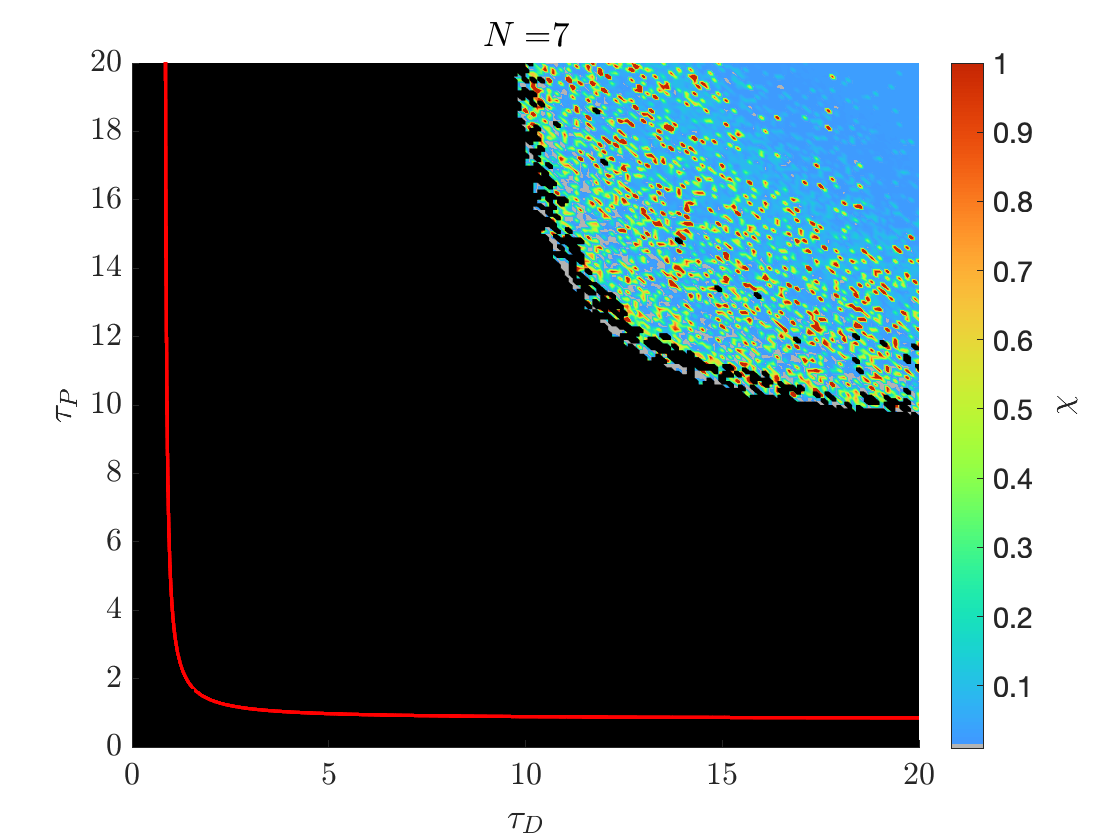}
    \caption{$P^{\text{ceil}}=0.8$}
    \end{subfigure}

  \begin{subfigure}[b]{0.4\textwidth}
    \includegraphics[width=\textwidth]{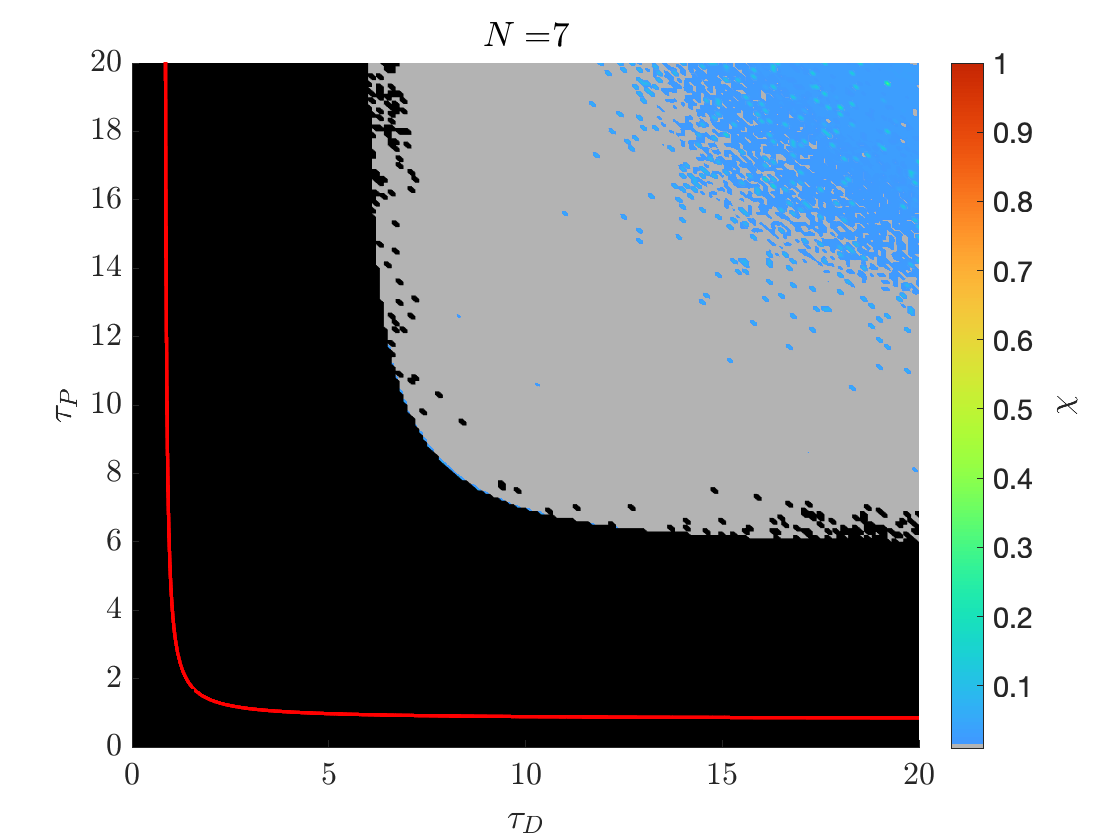}
    \caption{$P^{\text{ceil}}=0.85$}
    \end{subfigure}

\centering
    \begin{subfigure}[b]{0.4\textwidth}
    \includegraphics[width=\textwidth]{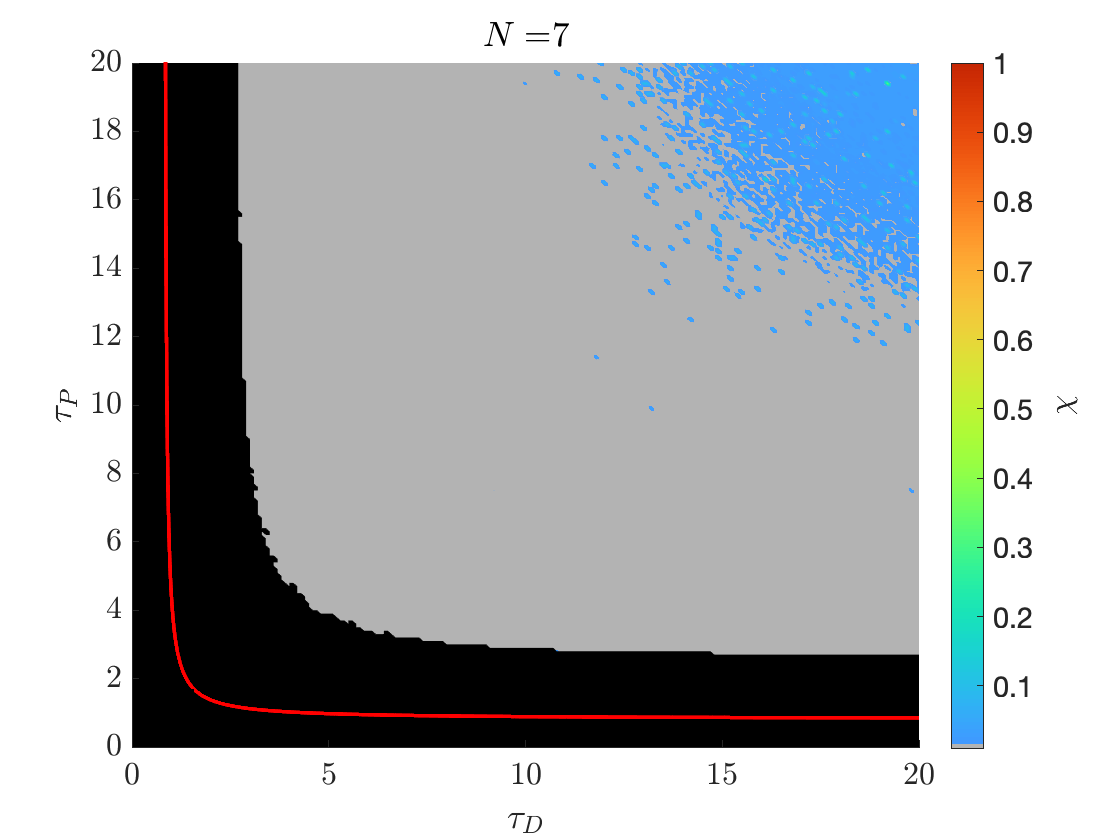}
    \caption{$P^{\text{ceil}}=0.9$}
    \end{subfigure}

  \begin{subfigure}[b]{0.4\textwidth}
    \includegraphics[width=\textwidth]{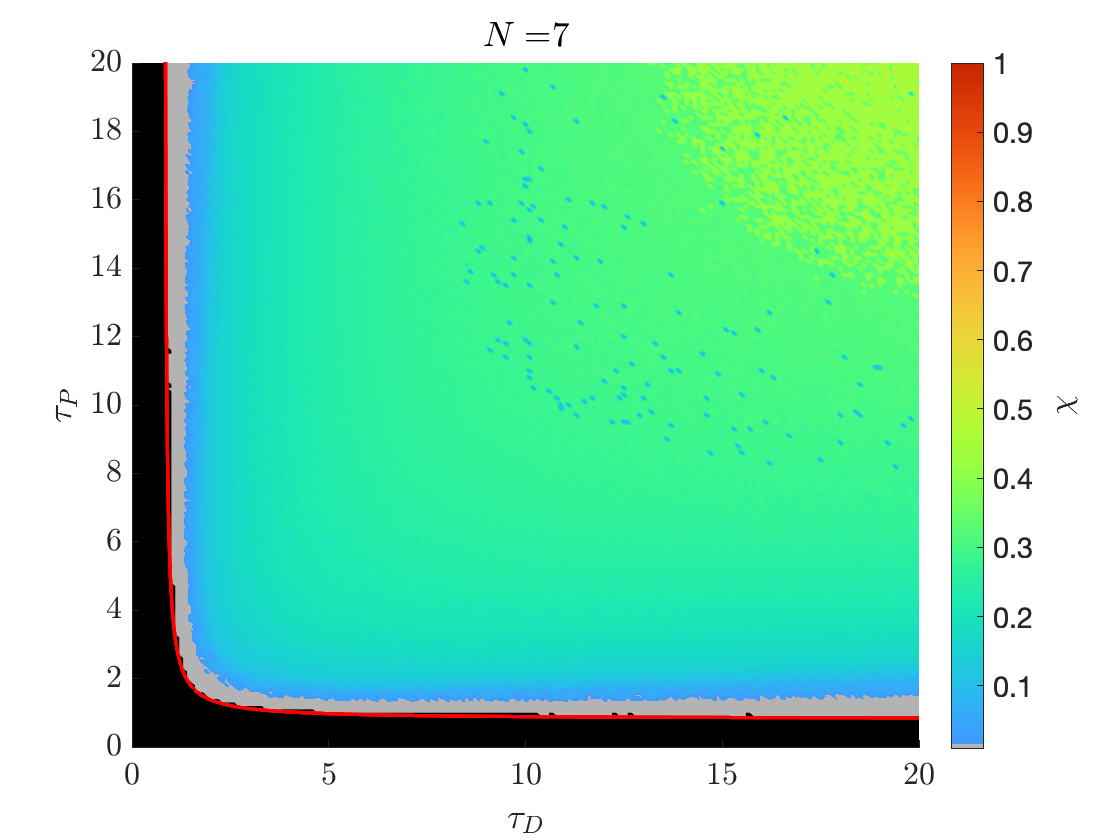}
    \caption{$P^{\text{ceil}}=0.95$}
    \end{subfigure}
    \caption{\justifying{$\chi$ as a function of $P^{\text{ceil}}$ in the $N=7$ case.}}
    \label{fig:RC chi influence}
\end{figure}
The specific values of $P_n$ at which the system settles are extremely sensitive to the initial conditions. To illustrate this, system \eqref{sys:model1} was solved numerically for $t \in [0,1000]$ with 10,000 sets of random initial conditions, with the same parameters as in Fig.~\ref{fig:Pequil values}. If an equilibrium solution was attained, as illustrated in Fig.~\ref{fig:Pequil values}, the equilibrium price values were stored. Figure \ref{fig:Pequil histogram} shows a histogram of the equilibrium values of $P_n$ collected from these simulations.

As one might expect, introducing a ceiling price that gives rise to infinitely many equilibrium solutions to system \eqref{sys:model1} increases the size of the parameter regime under which solutions to system \eqref{sys:model1} converge to an equilibrium, given $\mathcal{O}(1)$ random initial conditions. To explore this systematically, we fix $N = 7$ and plot  $\chi$ as a function of $\tau_P$ and $\tau_D$ in Fig.~\ref{fig:RC chi influence} for $P^{\text{ceil}}=0.8$ (a), $0.85$ (b), $0.9$ (c), and $0.95$ (d). As in Fig.~\ref{fig:chi maps}, a black color indicates that the prices settled at an equilibrium solution, and the red line indicates the onset of instability of the homogeneous equilibrium solution in the absence of rent control. With rent control there are regions in which the system converges at an equilibrium solution, while the equilibrium solution would be unstable without rent control (black regions above the red line). As $P^{\text{ceil}}$ approaches one, the behavior resembles the system without rent control, for which the prices are by construction bounded by one.
\begin{figure}[b]
    \centering
    \includegraphics[width = 0.4\textwidth]{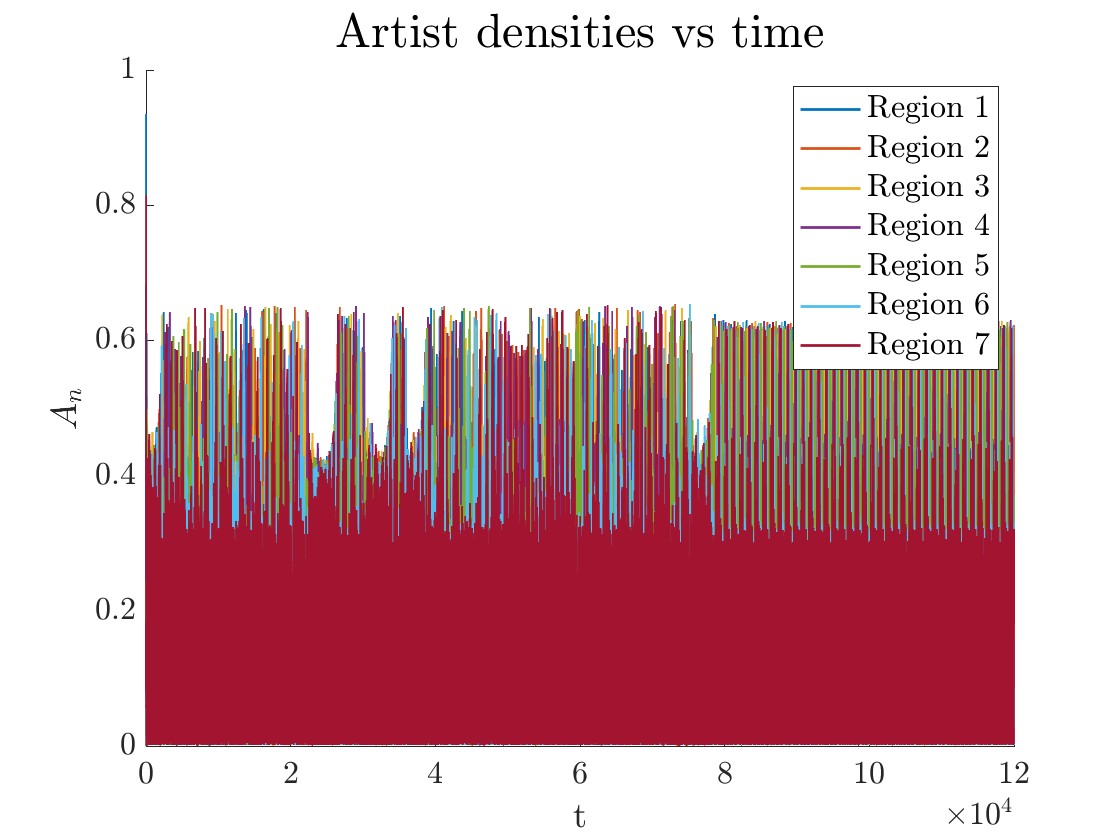}
    \caption{\justifying{Long-term transient chaos when $N=7, \ \tau_D = 18, \ \tau_P = 18,$ and $\Pceil = 0.79.$ }}
    \label{fig:transient chaos}
\end{figure}

While the introduction of rent control enlarges the regions where equilibrium solutions are stable, it can also result in more complex dynamics when equilibrium solutions are not reached. Specifically, some parameter choices seem to induce long-term transient chaos. For example, the artist populations as a function of time are shown in Fig.~\ref{fig:transient chaos} for $N = 7$, $\tau_P = \tau_D = 18$. An application of Gottwald’s 0-1 test to the time series of artists in region 1 within the time span $t \in [3050,69500]$ yields a $K$ value of approximately $0.9978$, strongly suggesting the time series within this range is (transiently) chaotic \cite{Skokos2016} (in Gottwald's 0-1 test, a $K$ value of 1 corresponds to chaos). 

The effects of a ceiling price are thus double-edged, as solutions to system \eqref{sys:model1} may be more disorganized than those to system \eqref{sys:model0}, but the parameter regime under which solutions converge to an equilibrium solution is expanded. 

\section{Discussion and Conclusions}\label{sec:discussionandconclusions}

In this study, we construct and analyze a simple dynamical system to better understand the influence of the so-called “creative class” \cite{florida2003cities} on the dynamics of gentrification. Current analytical theories on this phenomenon often consider binary demographic divisions and use agent-based modeling approaches \cite{benard2007wealth,zuk2015gentrification,laurie2003role,pancs2007schelling,zhang2004residential,bruch2006neighborhood}. While the model presented here is reductive, it captures the macroscopic temporal behavior of artists in an urban network undergoing rapid gentrification; namely, the continual displacement of the creative (or artist) class due to their influence on the cost of living in a neighborhood. System \eqref{sys:model0} models the interplay of the artist class, the desirability of a neighborhood, and the average price of real estate in a neighborhood in a simple fashion and admits desirable solutions, in the sense that equilibria exist and periodic or quasiperiodic displacement occurs under certain parameter regimes. 

The unique equilibrium solution to system \eqref{sys:model0}, given by the uniform distribution of artists, was found to be linearly stable as a function of parameters $N, \ \tau_D,$ and $\tau_P$, as prescribed by condition~\eqref{eq:stability condition}. Importantly, the linear stability of the spatially homogeneous solution does not imply system \eqref{sys:model0} is globally stable, as coexisting oscillatory solutions have been observed (see Fig.~\ref{fig:no convergence to the fixed point}). As whether system \eqref{sys:model0} admits chaotic time series is of interest in both a practical and a mathematical sense, we leverage the statistic $\chi$ to constrain the parameter regime under which we search for chaotic time series with significant amplitude variance. Gottwald's 0-1 test \cite{Skokos2016} was then applied to the time series of artists, and no chaotic behavior was found; we thus conjecture that system \eqref{sys:model0} does not admit chaotic solutions. 

System \eqref{sys:model1} is introduced to capture the influence of a rent control policy. While rent control policies are controversial, their effect on urban population dynamics is observed to be stabilizing in some cases \cite{Armsworth2005,gilderbloom2007thirty}, so the impact of a rent control policy on system \eqref{sys:model0} is of interest. Introducing a ceiling price $\Pceil$ such that the artists consider $\hat{P}=\min(P,\Pceil)$ in their decision-making gives rise to infinitely many equilibrium solutions. A numerical investigation of the time series generated by system \eqref{sys:model1} is conducted. As anticipated, the parameter regimes under which equilibrium solutions are reached are expanded as a function of $\Pceil$ (see Fig.~\ref{fig:RC chi influence}); namely, decreasing $\Pceil$ appears to have a stabilizing effect. Importantly, solutions to system \eqref{sys:model1} that do not converge to an equilibrium may be significantly more disorganized than those admitted by system \eqref{sys:model0}. This increased disorganization is clear in the values $\chi$ takes for the rent control system. Moreover, system \eqref{sys:model1} was found to admit long-term transient chaos under certain parameters (see Fig.~\ref{fig:transient chaos}); the effects of the implemented rent control policy on system \eqref{sys:model0} are thus double-edged. 

This work lends itself to expansion or modification in a large number of ways moving forward. One fundamental concern in the concept model shown in Fig.~\ref{fig:fc} is the omission of other significant drivers, such as a distinct minority class or entrepreneurs. While increasing the dimension of system \eqref{sys:model0} by introducing new time-dependent drivers is justified, we posit the problem will quickly become analytically intractable. Another future research direction is a spatial embedding of system \eqref{sys:model0}, which would lend itself to analysis via network theory tools. 
\\\\
\begin{acknowledgments}  Rodriguez and Shaw were partially funded by NSF-DMS-2042413 and AFOSR MURI FA9550-22-1-0380. Restrepo acknowledges support from NSF Grant No. DMS-2205967.
\end{acknowledgments}

\appendix 
\section{The unique equilibrium solution to system \eqref{sys:model0}}\label{Appendix A}

Direct substitution gives the spatially homogeneous solution: 
\begin{align}\label{eq:equil state vec}
   \mathbf{A^*} = 
    \begin{pmatrix}
    1/N\\
    1/N \\
    \vdots \\
    1/N
    \end{pmatrix}, \
    \mathbf{D^*} = \sigma(\mathbf{A^*}), \
    \mathbf{P^*} = \mathbf{D^*},
\end{align}
is an equilibrium solution of  \eqref{sys:model0}.

To prove this equilibrium solution is unique, consider ordering $P_n$ such that $P_1 \leq P_2 \leq \hdots \leq P_N$. If $P_N > P_n$ for any $n \in \{1,2,\hdots,N-1\}$ and $A_N \neq 0$, artists will move from region $N$ to region $n$, as described by system \eqref{sys:model0}, and it cannot be the case that $\frac{d A_N}{dt} = 0$. Furthermore, it cannot be the case that $A_N = 0$ if $P_N>P_n$ for some $n < N$, since \eqref{sys:model0} would give $P_N = \sigma(0) > P_n = \sigma(A_n)$, but since $\sigma(\cdot)$ is strictly monotonically increasing this would imply $0> A_n$.  Therefore, the unique equilibrium solution of system \eqref{sys:model0} is given by the uniform distribution of artists, $A_1 = A_2 = \hdots = A_N = 1/N$.

\section{Derivation of the linear stability condition \eqref{eq:stability condition}}\label{Appendix B}

Consider a perturbation to the homogeneous equilibrium solution
\begin{equation}
{\bf x^*}+{\bf {\delta x}} = 
    \begin{bmatrix}
        {\bf A^*} \\ 
        {\bf D^*} \\
        {\bf P^*}
    \end{bmatrix}
    +
        \begin{bmatrix}
        {\bf \delta A} \\ 
        {\bf \delta D} \\
        {\bf \delta P}
    \end{bmatrix}
\end{equation}
The linearized evolution of the perturbation satisfies
\begin{align}
\frac{d}{dt} {\bf {\delta x}} = M {\bf {\delta x}},
\end{align}
where the $3N \times 3N$ Jacobian matrix $M$ is given by the block matrix
\begin{align*}
    M = 
    \begin{pmatrix}
    M_{11} & M_{12} & M_{13} \\
    M_{21} & M_{22} & M_{23} \\
    M_{31} & M_{32} & M_{33}
    \end{pmatrix}, 
\end{align*}
where each block is an $N$-dimensional square matrix.

Blocks $M_{11}, \ M_{12},$ and $M_{13}$ correspond to the time rates of change of artists. Due to the discontinuous nature of the artist population evolution in \eqref{sys:model0}, we consider these blocks carefully.  After linearizing, the time derivative of each $\delta A_n$ takes the form 
\begin{align*}
    \tau_A \frac{d }{dt} \delta A_n = \sum_{k=1}^{N}\Bigl[ h(\delta P_k - \delta P_i)(A^*) - h(\delta P_i - \delta P_k)(A^*)\Bigr]
\end{align*}
The contribution  from the  $k^{th}$ region is 
\begin{align*}
    h(\delta P_k - \delta P_n)(A^*) - h(\delta P_i - \delta P_k)(A^*).
\end{align*}
Recalling that $h(x) = \max(x,0)$, we consider separately the cases $\delta P_k - \delta P_n \geq 0$ and $\delta P_k - \delta P_n < 0$, and find that in both cases we get $A^*(\delta P_k - \delta P_n)$. Therefore,
\begin{align*}
    \tau_A\frac{d}{dt} \delta A_n = \sum_{k=1}^{N}\Bigl[ A^*(\delta P_k - \delta P_n)\Bigr]
\end{align*}

Returning to the block matrix $M$,  for the state vector given in \eqref{eq:equil state vec}, $M_{11}$ and $M_{12}$ must be zero matrices, while $M_{13}$ must take the form
\begin{align*}
M_{13} = \frac{A^*}{\tau_A}
\begin{pmatrix}
(1-N) & 1 & \hdots & 1  \\[5pt]
1  & (1-N) & \hdots & 1  \\[5pt]
\vdots & \vdots & \ddots & \vdots \\[5pt]
1  & 1  & \hdots & (1-N)
\end{pmatrix}
.
\end{align*}
The entries of the remaining block matrices are simpler. Blocks $M_{21}, \ M_{22},$ and $M_{23}$ correspond to the time derivatives of desirability and take the forms
\begin{align*}
    \begin{cases}
    M_{21} = \frac{\sigma '(A^*)}{\tau_D} I, \\[5pt]
    M_{22} = -\frac{1}{\tau_D} I ,\\[5pt]
    M_{23} = 0,
    \end{cases}
\end{align*}
\noindent where $I$ represents the $n \times n$ identity matrix. Blocks $M_{31}, \ M_{32},$ and $M_{33}$ correspond to the time derivatives of price and take the forms
\begin{align*}
    \begin{cases}
    M_{31} = 0, \\[5pt]
    M_{32} = \frac{1}{\tau_P} I, \\[5pt]
    M_{33} = -\frac{1}{\tau_P} I.
    \end{cases}
\end{align*}
We then have
\begin{align} \label{M}
    M = 
    \begin{pmatrix}
      0 & 0 & M_{13} \\
      M_{21} & M_{22} & 0 \\
      0 & M_{32} & M_{33}
    \end{pmatrix}.
\end{align}
A straightforward application of the methods presented in Ref.~\cite{Powell2011} to calculate the eigenvalues of \eqref{M} yields
\begin{widetext}
\begin{equation}
    \det(M - \lambda I)= (-1)^{N}\lambda\left( \frac{1}{\tau_D} + \lambda \right) \left( \frac{1}{\tau_P} + \lambda \right) \left[\frac{\sigma'(1/N)}{\tau_D \tau_P }+\lambda\left(\frac{1}{\tau_D} + \lambda \right) \left(\frac{1}{\tau_P} + \lambda \right) \right]^{N-1}.
\end{equation}
\end{widetext}

Thus, three eigenvalues of $M$ have algebraic multiplicity 1 and are independent of $N$,
\begin{align*}
    \begin{cases}
    \lambda_1 = 0, \\
    \lambda_2 = -\frac{1}{\tau_D}, \\
    \lambda_3 = -\frac{1}{\tau_P},
    \end{cases}
\end{align*}
while the remaining eigenvalues are solutions to 
\begin{equation}\label{eq:eigenvalues}
  \frac{\sigma'(1/N)}{\tau_D \tau_P }+\lambda\left(\frac{1}{\tau_D} + \lambda \right) \left(\frac{1}{\tau_P} + \lambda \right) = 0.
\end{equation}
To identify the points where system \eqref{sys:model0}  becomes linearly unstable, we let $\lambda = i\omega$ in \eqref{eq:eigenvalues}, yielding

\begin{align*}
    \sigma'(A^*) = \omega \tau_A(1 + \omega \tau_P i)(\omega \tau_D -i)
\end{align*}
As the imaginary part of the right-hand side above must be equal to zero at the onset of instability, we obtain the linear stability condition (recalling that  we set $\tau_A = 1$)
\begin{align*}
    \sigma'(A^*) < \frac{1}{\tau_D} + \frac{1}{\tau_P}.
\end{align*}

\section{Definition of the parameter $\chi$} \label{Appendix C}

\begin{figure}[t]
    \centering
    \includegraphics[width = 0.4\textwidth]{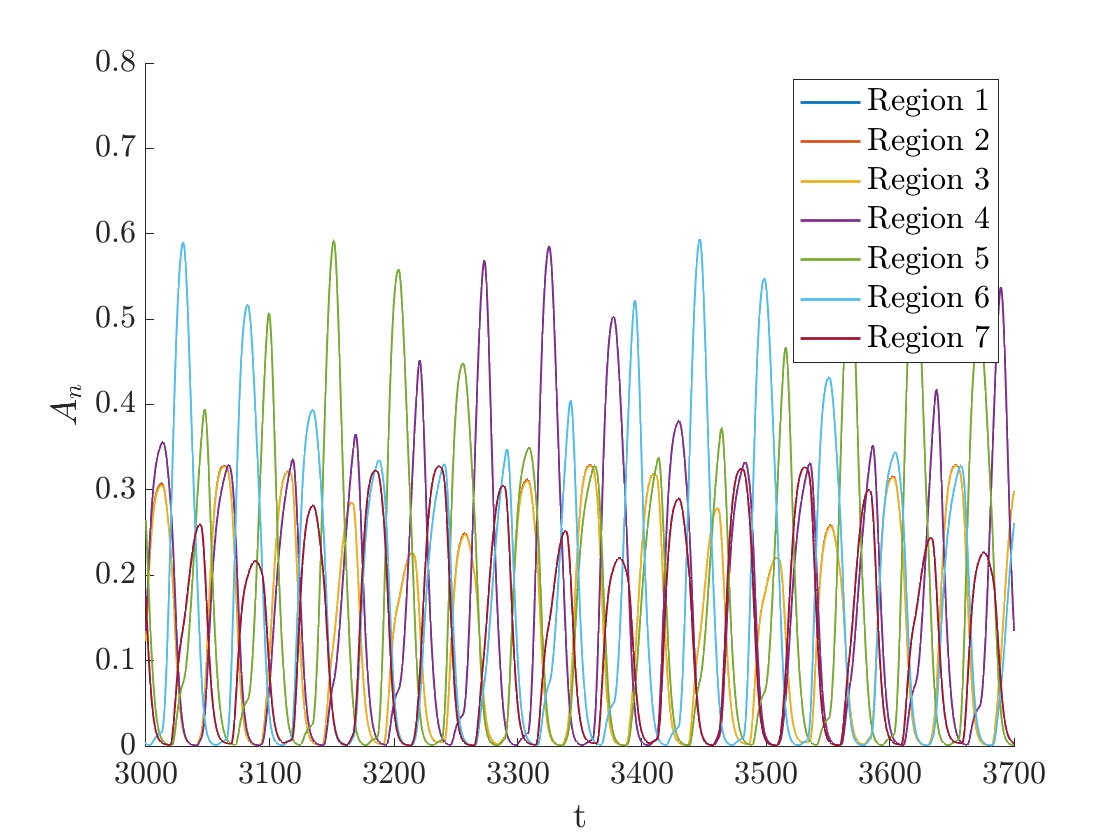}
    \caption{Disorganized oscillatory solution.}
    \label{fig:disorganized}
\end{figure}

\begin{figure}[t]
    \centering
    \includegraphics[width = 0.4\textwidth]{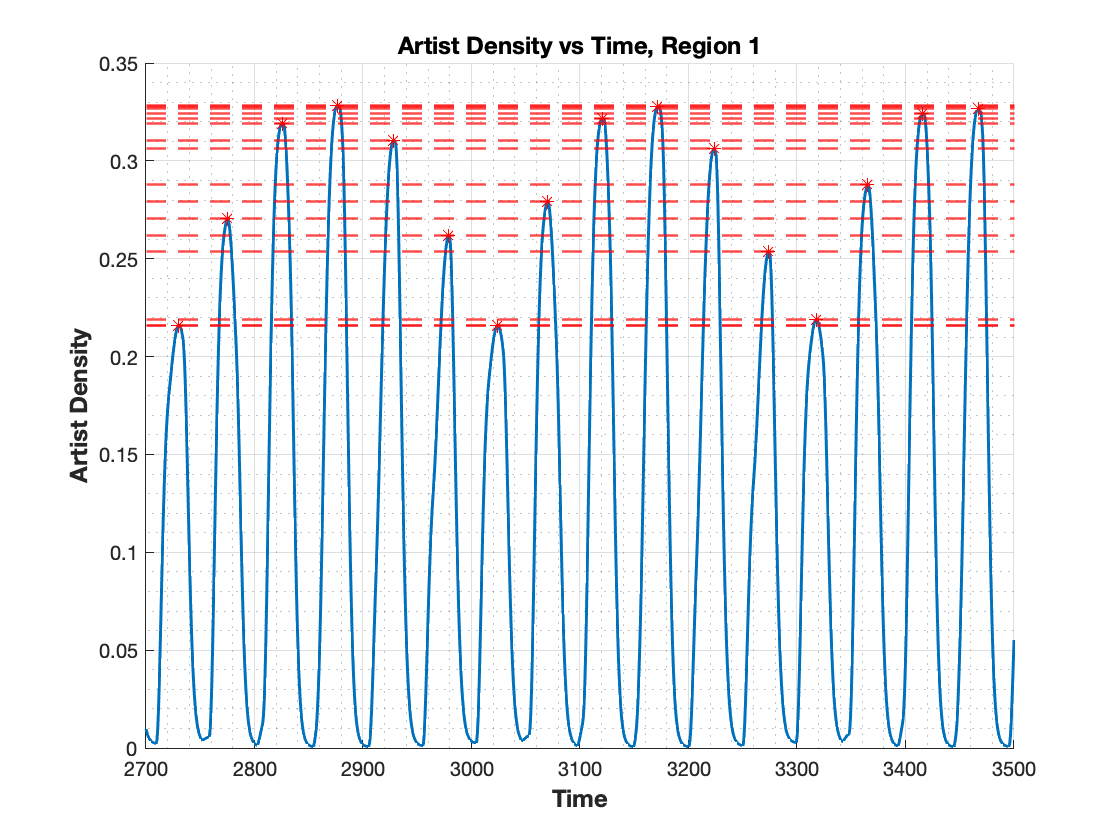}
    \caption{Marking the amplitudes of peaks in region 1.}
    \label{fig:amplitudes}
\end{figure}

Here we motivate the definition of the parameter $\chi$, which we used to quantify the degree of ``disorganization'' of solutions to \eqref{sys:model0}. By disorganization we mean a high degree of variability in the artist population oscillations, as illustrated in Fig.~\ref{fig:disorganized}.
To quantify this variability, given the time series of the artist populations $A_n$ in the $N$ regions, we define $\sigma_n$ to be the standard deviation of the local maxima of the artist population in the region $n$ (Fig.~\ref{fig:amplitudes} shows the local maxima of a particular time series as red-dashed lines). Similarly, we define $m_n$ to be the average value of the artist population in region $n$.

Then, we define $\chi$ to be 
\begin{equation}\label{eq:chi def}
    \chi  = \frac{2}{N}\sum_{n=1}^N\frac{\sigma_n}{m_n}.
\end{equation}
Given this definition, we have that $\chi \ll 1$ only if, on average, $\sigma_n \ll m_n$ (we note that, in practice, and as expected by symmetry, we have found that the values of $\sigma_n/m_n$ are similar for all $n$). That is, $\chi \ll 1$ if the variations in the local maxima are small compared with the size of the artist population oscillations. Cases with large variability give $\chi \approx 1$. 

\section{Marginal stability of rent-control equilibria}\label{sec:RC linear stability analysis}\label{Appendix E}

Consider the set $S = \{{\bf P} \in \R^N \ | \ P_n \in [P^{\text{ceil}}, 1] \ \forall n \in \{1,2,\hdots, N\}\}$. We show that equilibria of the form given in Eq.~\eqref{eq:rent control equilibrium} are linearly marginally stable as long as ${\bf P^*} \in \text{int}(S)$.

If $\bfP \in \inter (S)$ at an equilibrium solution, then for a small enough random perturbation $\delta \bfP$, it can be assumed that $\bfP + \delta \bfP \in S$ as well, as $\hat{P} = \min(\bfP + \delta \bfP,\Pceil) = \Pceil$. The perturbation matrix $M$ given by Eq.~\eqref{eq:M matrix} is transformed in this case to 
\begin{align} \label{Mat}
    M = 
    \begin{pmatrix}
      0 & 0 & 0 \\
      M_{21} & M_{22} & 0 \\
      0 & M_{32} & M_{33}
    \end{pmatrix},
\end{align}
where 
\begin{align}\label{eq:RC M matrix 1}
    M_{21} = 
    \begin{pmatrix}
     \sigma'(A_1) & 0 & \hdots & 0 \\
     0 & \sigma'(A_2) & \hdots & 0 \\
     \vdots & \ddots & \ddots & \vdots \\
     0 & 0 & \hdots & \sigma'(A_N)
    \end{pmatrix},
\end{align}
and 
\begin{figure}[b]
    \centering
    \includegraphics[width = 0.4\textwidth]{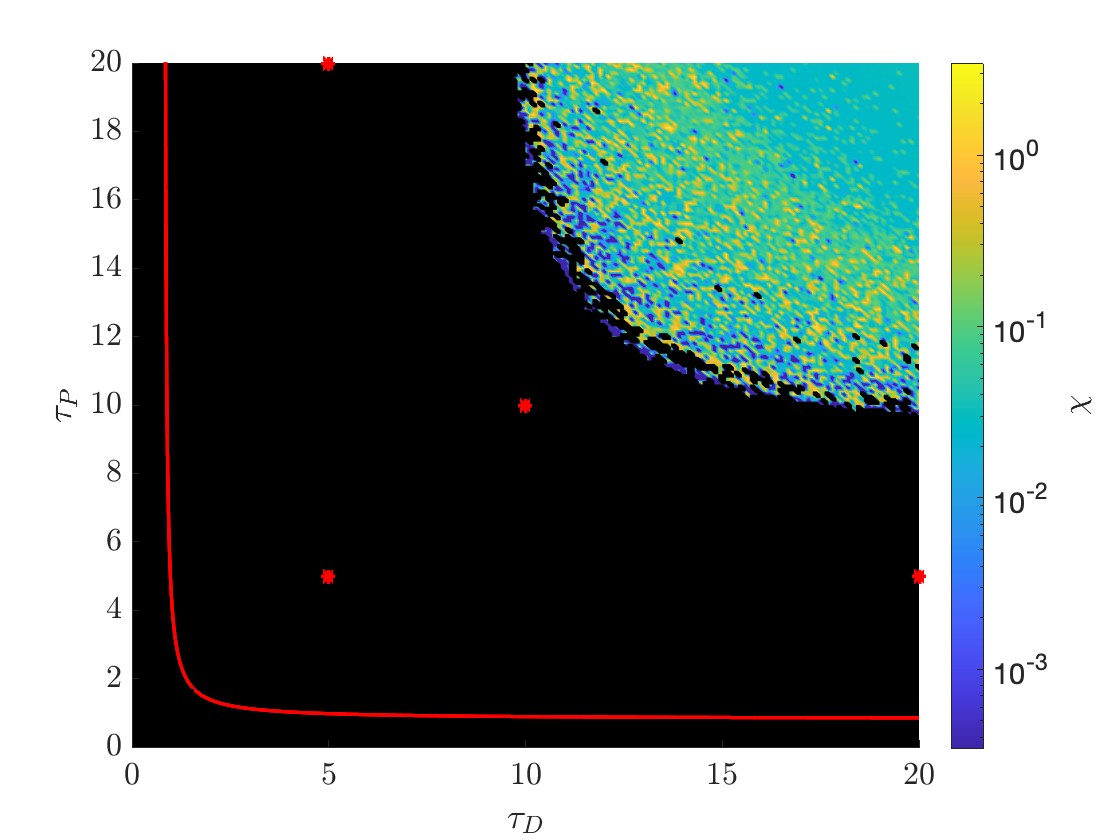}
    \caption{Parameter choices for the $N = 7$, $\Pceil = 0.8$ case.}
    \label{fig:neq7_Pceil_0pt8_hist}
\end{figure}
\begin{align}\label{eq:RC M matrix 2}
    \begin{cases}
    M_{22} = -\frac{1}{\tau_D} \mathbb{I} ,\\[5pt]
    M_{32} = \frac{1}{\tau_P} \mathbb{I}, \\[5pt]
    M_{33} = -\frac{1}{\tau_P} \mathbb{I}.
    \end{cases}
\end{align}
The eigenvalues of this matrix, each of algebraic multiplicity $N$, are given by 
\begin{equation*}
    \lambda_1 = 0, \ \lambda_2 = -\frac{1}{\tau_D}, \ \lambda_3 = -\frac{1}{\tau_P},
\end{equation*}
implying marginal stability of the equilibrium. 

The analysis above applies to equilibria such that ${\bf P} \in \inter (S)$. The following numerical experiments show that this is a generic behavior, and therefore we do not study the linear stability of equilibria such that ${\bf P} \in \partial S$. 

While a complete sweep of the parameter space $N, \tau_D, \tau_P$, and $\Pceil$ was not deemed feasible, a few parameter sets were used to generate histograms of equilibrium price values. For each $N \in \{1,2,\hdots,9\},$ $\Pceil$ was set to 0.8 and four points $(\tau_D,\tau_P) \in (0,20]\times(0,20]$ were chosen such that they were “far” from each other in the linearly stable parameter regime. An example selection of these points is shown in Fig.~\ref{fig:neq7_Pceil_0pt8_hist} for the $N=7$ case, where $(\tau_D,\tau_P) \in \{(5,5),(10,10),(5,20),(20,5)\}$ (red dots). 

As 40000 simulations were run for each $N \in \{1,2,\hdots,9\}$, approximately $1.56\times10^6$ equilibrium price values were calculated and no value was found to be equal to $\Pceil = 0.8$. We consequently conjecture that the boundary of the space $S$, as mentioned in Sec.~\ref{sec:RC linear stability analysis}, represents a theoretically possible but highly improbable equilibrium of system \eqref{sys:model1}; i.e. the probability that equilibrium solutions $\bfx^* = [{\bf A^*, \ D^*, \ P^*]^\mathrm{T}}$ of this system are attained such that ${\bf P^*}\in \partial S$ is approximately zero, so a linear stability analysis of this case is not necessary.

\bibliography{bibliography}

\end{document}